\documentclass[%
reprint,%
 amssymb, amsmath,%
 aip,pop,%
groupedaddress,%
]{revtex4-1}

\usepackage{color}
\usepackage{docs}%
\usepackage{bm}%
\expandafter\ifx\csname package@font\endcsname\relax\else
 \expandafter\expandafter
 \expandafter\usepackage
 \expandafter\expandafter
 \expandafter{\csname package@font\endcsname}%
\fi
\usepackage{graphicx}
\newcommand{\mnras}{MNRAS}
\newcommand{\aap}{A\&A}
\newcommand{\apj}{ApJ}
\newcommand{\polind}{\hat{\Gamma}}
\newcommand{\modind}{F}
\newcommand{\lrz}{\gamma}
\newcommand{\massint}{k}
\newcommand{\enth}{h}
\newcommand{\dens}{\rho}
\newcommand{\bflux}{A}
\newcommand{\vflux}{\Psi}
\newcommand{\mom}{P}
\newcommand{\belpot}{\chi}
\newcommand{\belpa}{\chi_A}

\newcommand{\ratb}{\zeta}

\newcommand{\gig}{f}

\newcommand{\nbl}{\nabla}

\begin{document}
\title{Rarefaction wave in relativistic steady magnetohydrodynamic flows}
\author{Konstantinos Sapountzis}\email{ksapountzis@phys.uoa.gr}
\author{Nektarios Vlahakis}\email{vlahakis@phys.uoa.gr}
\affiliation{
Faculty of Physics, University of Athens, 15784 Zografos, Athens, Greece}

\date{Received/Accepted}

\begin{abstract}
We construct and analyze a model of the relativistic steady-state magnetohydrodynamic (MHD) rarefaction that is induced when a planar symmetric flow (with one ignorable Cartesian coordinate) propagates under a steep drop of the external pressure profile. Using the method of self-similarity we derive a system of ordinary differential equations that describe the flow dynamics. In the specific limit of an initially homogeneous flow we also provide analytical results and accurate scaling laws. We consider that limit as a generalization of the previous Newtonian and hydrodynamic solutions already present in the literature. The model includes magnetic field and bulk flow speed having all components, whose role is explored with a parametric study.
\end{abstract}

\maketitle

\section{Introduction}
\label{introduction}

When a flow passes an acute vertex of an angle the information of the boundary's first derivative discontinuity propagates in the flow, and if the velocity of the flow exceed that of the fastest disturbances, i.e. the fast magnetosonic velocity, that propagation is performed in the form of a rarefaction wave; the resulting flow suffers a weak discontinuity\footnote{A discontinuity on the derivatives of the flow quantities rather than on the quantities itself, in our case first derivatives discontinuity.}. The importance of the rarefaction waves is significant in a number of phenomena and as such the reader can find relevant studies in various conditions and environments. The Newtonian hydrodynamic case is presented in many textbooks (see for example Landau \& Lifschitz\cite{Landau_fluid}), while a relativistic analytical approach under the same conditions was made by Granik\cite{Granik_1982}.

Moreover, the relativistic and highly magnetized counterpart of the phenomenon is mainly met at the high energy astrophysics where these extreme conditions apply, notably in Gamma-Ray Bursts.
Most astrophysical settings are considered as axisymmetric with ignorable azimuthal ($\phi$) coordinate,
in which case the study can be done on the so-called poloidal plane.
At sufficiently large cylindrical distances from the symmetry axis,
axisymmetry can be well approximated by planar symmetry, with the $\hat \phi$ direction replaced by the $\hat y$ direction in a Cartesian system of coordinates. We can continue to use the term ``poloidal plane'' for the $xz$ plane of this system and
split all vector quantities in ``poloidal'' (i.e., projections on the $xz$ plane)
and transverse ($\hat y$) components.
The interested reader is referred to \cite{Tchekhovskoy_Narayan_2010,Komissarov_Vlahakis_Konigl_2010} for numerical simulations and relevant discussions for axisymmetric and planar-symmetric Gamma-Ray Burst flows, \cite{Mizuno_2008, Zenitani_2010} for simulations including an external pressure profile, and \cite{Kostas_2013} for a semi-analytical planar-symmetric model containing only transverse magnetic field ($B_y$).

In this paper we present a general semi-analytical model which, besides $B_y$ contains a poloidal magnetic component of arbitrary magnitude, and discuss its potential implications. We also study the effect of the initial transverse velocity, and derive accurate scaling laws for the flow physical quantities. Our study is performed in the framework of the planar symmetric, ideal and relativistic steady-state MHD. The procedure is similar to the one followed in \cite{Kostas_2013}, but also to the hydrodynamic approach of \cite{Landau_fluid} and its relativistic counterpart \cite{Granik_1982}, and it is based on the class of the $r$ self-similar solutions. Beyond its potential astrophysical applications, the theoretical importance of rarefaction is evident, and the aim of the present work is to provide an insight to the relativistic magnetized regime, and thus to serve as a generalization of the already available hydrodynamical solutions.

We use the similarity property to degrade the system of the high nonlinear partial differential equations to a system of ordinary differential ones that are easier to manipulate. In section~\ref{seceq} we present the full steady-state equations, and in section~\ref{secsmodel} we apply the self-similarity to obtain the semi-analytical system. In section~\ref{results} we integrate the resulting system using a simple numerical algorithm for cases where the relative significance of the poloidal magnetic field alters. We also included some models with different initial transverse velocities, as also three models corresponding to the numerical simulations of \cite{Mizuno_2008}, in order to check further the validity of our model. Section \ref{discussion} contains the relevant discussion, while in Appendix~\ref{appA} we derive analytical scaling laws for the interesting case of a cold and homogeneous flow.

\section{Steady-state equations}
\label{seceq}

The system of relativistic MHD equations is expressed by the equations determining the hydrodynamical properties of the flow under the influence of the electromagnetic field (emf), Maxwell's and Ohm's laws. The energy-momentum tensor is constructed as the superposition of the matter ($T^{\mu\nu}_{\rm{hy}}$) and the emf ($T^{\mu \nu}_{\rm{EM}}$) tensors
\begin{equation}
T^{\mu \nu}=T^{\mu \nu}_{\rm{hy}}+T^{\mu \nu}_{\rm{EM}} \,.
\end{equation}
The former one is given by
\begin{equation}
T^{\mu \nu}_{\rm{hy}}=\enth \dens u^{\mu} u^{\nu} +p n^{\mu \nu} \,,
\end{equation}
where $u^{\nu}=\left(\lrz c,\lrz \bm{v}\right)$ the plasma four-velocity, $\bm{v}$ the three-velocity and $\lrz=1/(1-v^2/c^2)^{1/2}$ the Lorentz factor. Neglecting gravity and general relativistic effects we chose a Minkowski metric $g^{\mu\nu}=n^{\mu\nu}=\left(-,+,+,+\right)$. The thermodynamical parameters $h$, $p$, and $\dens$ are the enthalpy per rest energy, pressure, and matter density of the plasma as measured in the comoving frame. For a gas obeying a polytropic equation of state the following relation applies
\begin{equation}
    \label{eqenth0}
    \enth =1+\frac{\polind}{\polind-1}\frac{p}{\dens c^2}
\end{equation}
with $\polind$ the usual polytropic index.

The components of the $T^{\mu \nu}_{\rm{EM}}$ in analytical form are
\begin{equation}
T^{00}_{\rm EM}=\frac{E^2+B^2}{8 \pi} \,, \quad T^{0j}_{\rm EM}=T^{j0}_{\rm EM}=\left(\frac{\bm{E} \times {\bm{B}}}{4 \pi} \right)_j \,,
\end{equation}
\begin{equation}
T^{jk}_{\rm EM}=-\frac{E_j E_k + B_j B_k}{4 \pi}+\frac{ E^2 + B^2}{8 \pi} \eta^{jk} \,,
\end{equation}
where Latin indices $i,j=1,2,3$ stand for the spatial coordinates, while Greek for all, and $\bm{E}, \bm{B}$ the electric and magnetic field as measured in the laboratory frame. We can identify $T^{00}_{\rm EM}\,, T^{j0}_{\rm EM}\,, T^{jk}_{\rm EM}$ as emf energy density, energy flux, and magnetic stress contributions, respectively. The full energy-momentum tensor provides the equations of motion in the covariant form $T^{\mu\nu}_{,\nu}=0$, but a straightforward use of all of these equations leads to difficult manipulating forms. Thus it is a common practice to substitute some of them with other equivalent, but simpler ones, as explained below.

At the steady-state limit, the continuity equation $(\dens u^{\nu})_{\,,\nu}=0$, is written in vector form
\begin{equation}
\label{continuity}
\nbl \cdot \left( \lrz \dens\bm{v} \right) =0 \,.
\end{equation}
The projection of the energy-momentum equation on the proper time direction ($u_\nu T^{\mu\nu}_{,\nu}=0$) provides the entropy conservation
\begin{equation}
\label{entropycons}
 \bm{v} \cdot \nbl{ \left( \frac{p}{\dens^{\polind}}\right) } = 0 \,.
\end{equation}
The polytropic index takes the adiabatic values, $4/3$ and $5/3$ in the limits of ultrarelativistic and nonrelativistic temperatures, respectively.

Maxwell's equations for the steady-state become
\begin{eqnarray}
\label{Maxwell}
{\nbl} \cdot {\bm{B}}=0 \,, \quad \nbl \cdot {\bm{E}}= \frac{4 \pi}{c} J^0 \,,
\nonumber\\
\nbl \times {\bm{B}}= \frac{4 \pi}{c}{\bm{J}}\,, \quad \nbl \times {\bm{E}}=0\,,
\end{eqnarray}
where $J^{\nu}=\left(J^0,\bm{J}\right)$ the four-current, $J^0/c$ the charge density and $\bm{J}$ the current density. Moreover, at the limit of an infinite electrical conducting plasma the comoving electric field is zero, and Ohm's law yields
\begin{equation}
\label{ohms}
{\bm{E}}=-\frac{\bm{v}}{c} \times\bm{B}\,.
\end{equation}
We can write explicitly the spatial components of the momentum equation using Maxwell's equations as
\begin{equation}
\label{spatmomentum}
-\lrz \dens \left({\bm{v}} \cdot \nbl \right) \left(\enth \lrz {\bm{v}} \right)
-\nbl p +
\frac{J^0 {\bm{E}}+{\bm{J}} \times {\bm{B}}}{c} =0 \,.
\end{equation}
Equations (\ref{continuity}--\ref{spatmomentum}) together with the boundary conditions determine in principle the flow, but the high nonlinear character make this task rather difficult.

\begin{figure}
    \includegraphics[width=0.42\textwidth,angle=0]{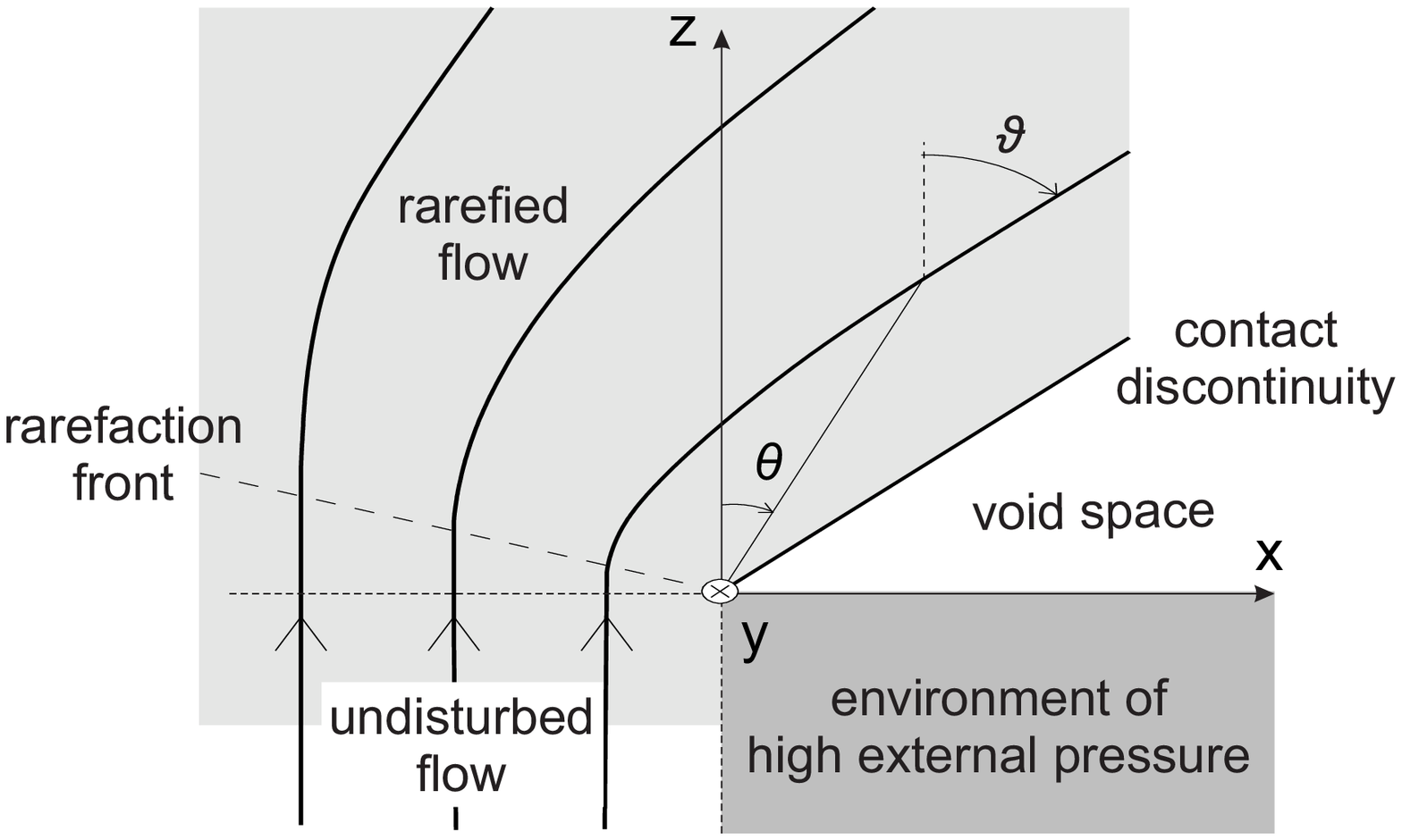}
    \caption{The geometry of a planar symmetric rarefied flow and the coordinate system.
The coordinate $y$ is ignorable and the plane $xz$ is the ``poloidal'' plane.
Notice the three regions that in principle exist: the undisturbed plasma
(which is in pressure equilibrium with its environment),
the rarefied one, and the vacuum.
The situation is similar to a supersonic hydrodynamic flow around an acute angle;
here the flow is magnetized and it is super-fast magnetosonic.
The weak discontinuity, i.e. the rarefaction front, and the contact discontinuity separating the plasma fluid from the void space, are also shown. Angles $\theta$, $\vartheta$ stand for the polar angle and the poloidal field/streamline inclination respectively; both are measured from the $z$-axis clockwise.}
\label{geom}
\end{figure}

We carry a first partial integration assuming Cartesian coordinates with the axis origin on the boundary discontinuity and planar symmetry along the $\hat y$ direction $\left(\partial / \partial y=0\right)$, see Fig.~\ref{geom}.
From Faraday's law the electric field is related to an electric potential, $\bm E=-\nbl V$. Assuming that the potential is also planar symmetric, $V=V(x,z)$, we find that the transverse electric field vanishes ($E_y$=0). That symmetry in conjunction with Ohm's law provides $\bm{v}_p \parallel \bm{B}_p$, and we can therefore write the flow velocity in the form
\begin{equation}
\label{masintdef}
    {\bm{v}}=\frac{\massint}{\lrz \dens}{\bm{B}} + c\belpot \hat y \,,
\end{equation}
where
\begin{equation}
\massint =\lrz \dens \frac{v_p}{B_p} \,, \quad
    \belpot =\frac{v_y}{c}-\frac{v_p}{c}\frac{B_y}{B_p} \,.
\end{equation}
Both quantities $\massint$, $\belpot$ are integrals of motion, i.e. remain constant along a poloidal streamline (or field line). The former integral stands for the ratio of the mass to the magnetic flux, while the second one is investigated later. Furthermore, we introduce the fluxes per unit length in the $\hat y$ direction, $\bflux$ and $\vflux$, to label the poloidal magnetic field lines and the poloidal streamlines, respectively:
\begin{equation}
    \bflux =\int \bm B_p \cdot d\bm s\times\hat y \,, \quad \vflux=\int\lrz \dens \bm v_p \cdot d\bm s\times\hat y \,,
\end{equation}
where the integration is performed on a line on the polidal plane, starting from a point of the $z$ axis. Reverting the above relationships we obtain
\begin{equation}
    \bm B_p= \nbl\bflux \times \hat y \,, \quad \bm v_p=\frac{1}{\lrz\dens} \nbl\vflux \times \hat y \,,
\end{equation}
where the equivalency of whether we use the poloidal magnetic field lines, or the poloidal streamlines is stated explicitly, $\nbl \vflux=\massint \nbl\bflux$. In this paper we use the magnetic field line notation to project the energy momentum equation parallel ($\hat{b}=\bm{B}_p / B_p$) and perpendicularly ($\hat{n}=-\nbl \bflux / \left|\nbl \bflux\right|$) to the field lines direction; notice that in the limit of the negligible poloidal magnetic field this choice poses some easy-lifted complications of interpretation nature, see the end of next section.

The constancy of $\massint$ is derived by inserting the velocity form (\ref{masintdef}) in the continuity equation~(\ref{continuity}), using the planar symmetry and the zero divergence of $\bm B$ to obtain $\bm B_p \cdot \nbl\massint =0 \Leftrightarrow\massint=\massint(\bflux) $. The substitution of the same velocity expression in Ohm's law (\ref{ohms}) yields
\begin{equation}
\label{electrictoBp}
    {\bm{E}}=-\belpot \nbl \bflux \,, \quad E=\belpot B_p \,,
\end{equation}
which in conjunction with the Faraday's law provides also the constancy of $\belpot$. It is useful to compare that integral with the corresponding Ferraro's isorotation law in axisymmetric flows, which is related to the so-called light cylinder; see for example \cite{beskin_mhd} for a general analysis.
On this cylinder the ratio $E/B_p$, which in the axisymmetric case is a function of the cylindrical distance, becomes unity. In the planar symmetric case no such cylinder exists, and this is reflected to the constancy of the ratio $E/B_p$. This is going to have an important role to the power laws derived later.

Moreover, Eq.~(\ref{entropycons}) provides the usual polytropic equation
\begin{equation}
\label{eqQdef}
    Q(\bflux)=p / \dens^{\polind} \,,
\end{equation}
and thus $Q(\bflux)$ integral states the entropy conservation along streamlines. Two more quantities complete the set of the integrals
\begin{eqnarray}
    \mom=\mom(\bflux) = \enth \lrz v_y -\frac{B_y}{4 \pi \massint} \,, \\
    \mu=\mu(\bflux)= \enth \lrz - \frac{\belpot B_y}{4 \pi \massint c} \,,
\end{eqnarray}
and stand for the total (matter + emf) momentum-to-mass flux ratio, and the
total energy-to-mass flux ratio, respectively. No more independent integrals exist, but a useful combination which appears often in the subsequent calculations is $\belpa^2=\mom \belpot/(\mu c)$.\footnote{The notation was chosen in accordance with the axisymmetric flows in which $\belpa$ is the value of $\belpot$ at the Alfv\'enic surface, while the integral $\mom$ corresponds to the angular momentum-to-mass flux ratio.}

Besides the five integrals ($\massint$, $\belpot$, $\mu$, $\belpa$ or $\mom$, $Q$) two more equations are needed to fully determine the flow. For convenience we introduce the ``Alfv\'enic'' Mach number
\begin{equation}
\label{Mdef}
    M^2\equiv\frac{\left(\lrz v_p\right)^2}{B_p^2/(4 \pi \dens \enth)}=\frac{ 4 \pi \enth\massint^2}{\dens} \,,
\end{equation}
and the magnetization parameter
\begin{equation}
\label{sigdef}
\sigma=-\frac{\belpot B_y}{4 \pi \lrz \enth c \massint} \,,
\end{equation}
i.e., the ratio of the Poynting to mass energy flux. In terms of the above quantities the physical quantities are written as
\begin{eqnarray}
\label{primedef}
\dens=\frac{4 \pi \enth \massint^2}{M^2} \,,
\quad p=Q\dens^{\polind} \,,
\quad \enth= 1+\frac{\polind}{\polind-1}\frac{p}{\dens c^2} \,, \quad \\
\label{eqnprimedefB}
{\bm {B}} = \nbl \bflux \times \hat y -\frac{4 \pi \mu\massint c (\belpot^2 -\belpa^2)}{\belpot (M^2 +\belpot^2-1)} \hat y \,, \quad
{\bm{E}}=-\belpot \nbl \bflux \,, \quad \\
\lrz=\frac{\mu}{\enth} \frac{M^2+\belpa^2-1}{M^2+\belpot^2-1}\,, \quad \\
\lrz \frac{\bm{v}}{c}=\frac{M^2}{4 \pi c \massint \enth}
\nbl \bflux \times \hat y
+\frac{\belpa^2\mu}{\belpot\enth} \frac{M^2+\belpot^2-\belpot^2/ \belpa^2}{M^2+\belpot^2-1}\hat y \,. \quad
\end{eqnarray}

An interesting situation arises when the $M^2+\belpot^2-1$ denominator vanishes corresponding to the so-called Alfv\'enic surface. The requirement that $B_y$ remains finite at that surface yields $\belpot^2=\belpa^2$. Since these are integrals of motion they remain equal everywhere, meaning that $B_y=0$ and the flow carries no Poynting flux. For this reason magnetized planar symmetric flows cannot be trans-Alfv\'enic.

The initial conditions determine the integrals of motion, but one seeks for the quantities $\bflux$ and $M$ or $\enth$; the last two are related by the expression, using Eqs.~(\ref{primedef}):
\begin{equation}
\label{mach-h}
M^2=4\pi \massint^2\left(\frac{\polind}{\polind-1} \frac{Q}{c^2}\right)^{\frac{1}{\polind-1}} \enth \left(\enth-1\right)^{-\frac{1}{\polind-1}} \,.
\end{equation}

The two remaining equations are the Bernoulli (or wind equation)
\begin{eqnarray}
\label{bernoulli1}
\frac{\mu^2}{\enth^2} \frac{\left(M^2+\belpa^2-1\right)^2 - \left(\belpa^2/\belpot \right)^2 \left(M^2+\belpot^2-\belpot^2/\belpa^2\right)^2 }{\left(M^2 +\belpot^2-1\right)^2} \nonumber\\
=1 + \left(\frac{M^2 \nbl \bflux}{4 \pi c \massint \enth} \right)^2 \,, \qquad
\end{eqnarray}
which is obtained by substituting all the quantities in the identity
$\lrz^2 - (\lrz {v_y}/c)^2 =1+(\lrz {v_p}/c)^2$,
and the transfield equation obtained by projecting the momentum equation perpendicular to the magnetic field
\begin{eqnarray}
\label{transfield}
M^2 \mid \nbl \bflux \mid^2 \left[
\nbl^2 \bflux - \nbl \bflux \cdot \nbl \ln \mid \nbl \bflux \mid \right]
\nonumber  \\
-\frac{\polind-1}{\polind} \nbl \left[
16 \pi^2 \massint^2 c^2 \frac{\enth (\enth-1)}{M^2}
\right] \cdot \nbl \bflux
\nonumber  \\
+\belpot \frac{d\belpot}{d\bflux} \mid \nbl \bflux \mid^4
+\left( \belpot^2-1 \right) \nbl^2 \bflux  \mid \nbl \bflux \mid^2
\nonumber  \\
-\frac{1}{2} \nbl \left( \frac{4 \pi \massint \mu c}{\belpot}
\frac{\belpot^2 -\belpa^2}{M^2 +\belpot^2-1}
\right)^2 \cdot \nbl \bflux =0 \,.
\end{eqnarray}

Roughly speaking we can state that the solution of the transfield equation determines the shape of the streamlines, while Bernoulli the energetics along them, but this distinction is not clear neither fruitful. Both equations must be solved simultaneously and a simple inspection shows the difficulties involved. The task of finding an analytical solution in the general case seems impossible, and thus all the efforts are focused on the quest of a solution with specific symmetries suitable to describe the particular problem; in our case this is the self-similar shape of the poloidal streamlines.

\section{The \lowercase{$r$} self-similar model}
\label{secsmodel}
In order to induce the similarity property we assume that all the quantities have a dependence of the form $r^{\modind_i}f_i(\theta)$ where $r=\sqrt{x^2+z^2}$ the distance from the corner and $\theta$ the angle measured from the $z$-axis ($x=r\sin\theta$, $z=r\cos\theta$). Our goal is to determine the various exponents $\modind_i$ in such a way that the resulting differential equations will be separable on the variables $r$, $\theta$. The method of similarity is quite familiar and has been used before in a number of studies both in Newtonian \cite{Blandford_1982}, \cite{Vlahakis_Tsinganos_1998} or in the relativistic context \cite{Begelman_1992, Contopoulos_1994, Vlahakis_2003a}.

The substitution of the similarity expressions in Eqs.~(\ref{bernoulli1}, \ref{transfield}) is straightforward and by inspection we conclude that our model derives separable equations under the following forms
\begin{eqnarray}
\bflux=-r^\modind a(\theta) \,, \quad
\massint=r^Y \kappa(\theta) \,,  \quad
Q=r^Z q(\theta) \,,
\nonumber  \\
M=M(\theta) \,, \quad \enth=\enth(\theta) \,, \quad
\belpa^2\,,\mu\,,\belpot=\mbox{const}\,.
\end{eqnarray}
Substituting the above forms into the Bernoulli equation~(\ref{bernoulli1}), we obtain $\modind-Y-1=0$, while from Eq.~(\ref{mach-h}) we conclude that $Q$ follows the dependence $Z=-2\left(\modind-1\right)\left(\polind-1\right)$; notice that both $\massint$, $Q$ are integrals and thus their angular dependence is related to the one of the poloidal flux: $\kappa=\massint_0\, a^{1-1/\modind}$, $q = q_0\,a^{-2\left(\polind-1\right)\left(1-1/\modind\right)}$, with constant $\massint_0$ and $q_0$. For purely algebraic reasons, we use
$f(\theta)\equiv 4 \pi c k_0 /(\modind \belpot^2 a^{1/\modind})$ instead of $a$. Accordingly $f$ is proportional to the radial distance along the same magnetic field line (or streamline): for a line passing through ($r_0$, $\theta_0$) any other point obeys $r/r_0=f/f_0$ with $f_0=f(\theta_0)$. Moreover we introduce one more angle $\vartheta$ that stands for the angle between the poloidal magnetic field line (or streamline) and the $z$-axis: $\tan\vartheta=B_x/B_z$. Using the latter variable, we express the parallel $\hat b \equiv \bm B_p / B_p$ and the perpendicular direction $\hat n\equiv-\nbl \bflux / \left|\nbl \bflux \right|$ to the poloidal magnetic field lines as
\begin{eqnarray}
\hat b=\cos\left(\vartheta-\theta\right)\hat r+\sin\left(\vartheta-\theta\right) \hat\theta \,, \nonumber \\
\hat n=\sin\left(\vartheta-\theta\right) \hat r-\cos\left(\vartheta-\theta\right) \hat\theta \,.
\label{unitvecs}
\end{eqnarray}

Under these assumptions the expressions for the physical quantities become
\begin{eqnarray}
\label{eqnsdens}
\bflux=-\left(\frac{4\pi c \massint_0 r}{\modind \belpot^2 f}\right)^\modind \,,  \quad
\dens=\frac{4 \pi \enth \massint_0^2}{M^2}
\bflux^{2(\modind-1)/\modind}
\,, \quad \\
\label{eqnsby}
{\bm B}_p = \frac{-\modind \bflux}{r \sin\left(\vartheta-\theta\right)} \hat b \,,  \quad
{\bm B}_y=
\frac{\modind A \mu f \belpot (\belpot^2 -\belpa^2)}{r(M^2+ \belpot^2-1)} \hat y \,,\quad \\
\bm{E}=\frac{-\modind \bflux \belpot}{r \sin\left(\vartheta-\theta\right)} \hat n \,, \quad \\
\label{eqnsgamma}
\lrz =\frac{\mu}{\enth}\frac{M^2 + \belpa^2-1}{M^2+\belpot ^2-1} =\frac{\mu}{\enth\left( {1+\sigma}\right)}\,, \quad \\
\label{eqnsgamvp}
\frac{\lrz \bm v_p}{c}=\frac{M^2}{\belpot ^2 f \enth\sin\left(\vartheta-\theta\right)}\hat b \,, \quad \\
\label{eqnsgamvy}
\frac{\lrz\bm v_y}{c}=\frac{\belpa^2 \mu}{\enth \belpot}\frac{M^2+\belpot ^2 - \belpot^2/ \belpa^2}{M^2+\belpot^2-1}\hat y \,. \quad
\end{eqnarray}

Before proceeding further to the equations involved, we note that the angle $\vartheta$ is related to the derivative of the function $f$:
The form of $A\propto (r/f)^F$ implies that $\nbl A$ is parallel to $f\hat r - (df/d\theta) \hat \theta$,
and since $\hat n=-\nbl A / |\nbl A| $ a first equation is obtained using Eq.~(\ref{unitvecs})
\begin{equation}
\label{feq}
\frac{df}{d\theta}=\frac{f}{\tan\left(\vartheta-\theta\right)} \,.
\end{equation}
The Bernoulli equation~(\ref{bernoulli1}) is now written as
\begin{eqnarray}
\label{bernoulli}
\frac{\mu^2}{\enth^2} \frac{\left(M^2+\belpa^2-1\right)^2 - \left(\belpa^4/\belpot^2 \right) \left(M^2+\belpot^2-\belpot^2 / \belpa^2\right)^2 }{\left(M^2 +\belpot^2-1\right)^2} \nonumber\\
=1 + \left[\frac{M^2}{\belpot^2 \enth f \sin\left(\vartheta-\theta\right)} \right]^2 \,. \qquad
\end{eqnarray}
Besides its algebraic form the differential one is also used at the subsequent calculations:
\begin{eqnarray}
\label{eqvartheta}
\frac{1}{\tan\left(\vartheta-\theta\right)} \frac{d\vartheta }{d\theta}
=\frac{\belpot ^4 f^2 \enth\sin^2\left(\vartheta-\theta\right)}{M^4}\frac{d\enth}{d\theta} \nonumber \quad \\
+\left[1-\frac{\belpot^2 \mu^2 f^2\left(\belpot^2 - \belpa^2 \right)^2 \sin ^2\left(\vartheta-\theta \right)}{\left(M^2+\belpot ^2-1\right)^3}\right]\frac{1}{M^2}\frac{dM^2}{d\theta} \,. \quad
\end{eqnarray}
Equation~(\ref{mach-h}) provides a relationship between $M^2$ and $\enth$
(both are functions of $\theta$ alone; note the  $Q$ and $\massint$ are constants along streamlines
and their combination $Q k^{2(\polind-1)} $ is a global constant), implying
\begin{eqnarray}
\label{eqenth}
\frac{d\enth}{d\theta}=-\frac{\enth u_s^2}{M^2}\frac{d M^2}{d\theta}\,, \quad u_s^2=\frac{\left(\polind-1\right)\left(\enth-1\right)}{\polind-1+\left(2-\polind\right)\enth} \,,
\end{eqnarray}
where $u_s=c_s/\sqrt{c^2-c_s^2}$ the sound proper velocity (over $c$), with $c_s=\sqrt{\polind p / (\dens \enth)}$.

Applying the similarity expressions to the transfield equation~(\ref{transfield}) we find
\begin{eqnarray}
(\modind-1)\left[
\frac{\mu^2(\belpot^2-\belpa^2)^2}{(M^2 +\belpot^2-1)^2}+
\frac{2(\polind-1) \enth (\enth-1)\belpot^2}{\polind M^2}\right]
=
\nonumber \quad \\
\frac{1}{\belpot^2 \gig^2}\left[\frac{\belpot^2-1}{\sin^2(\vartheta-\theta)}+M^2\right]
\frac{d\vartheta}{d\theta}
+\frac{(\modind-1)(\belpot^2-1)}{\belpot^2 \gig^2\sin^2(\vartheta-\theta)}
\nonumber \qquad \\
+\frac{\sin(2\vartheta-2\theta)}{2}
\left[
\frac{\belpot^2\enth}{M^2}\frac{d\enth}{d\theta}
-\frac{\mu^2(\belpot^2-\belpa^2)^2}{(M^2 +\belpot^2-1)^3}
\frac{dM^2}{d\theta}
\right]
\,, \qquad
\end{eqnarray}
which, in combination with Eqs.~(\ref{eqvartheta}, \ref{eqenth}) gives an equation for $M$:
\begin{equation}
\label{Meq}
\frac{dM^2}{d\theta}=\frac{\left(\modind-1\right) M^2}{\tan\left(\vartheta-\theta\right)}\frac{N}{D} \,,
\end{equation}
\begin{eqnarray*}
N=-\frac{M^2\left(\belpot^2-1\right)}{\belpot^4 f^2 \enth^2 \sin^2\left(\vartheta-\theta\right)}
+\frac{2\left(\polind-1\right)\left(\enth-1\right)}{\polind \enth}
\nonumber\\
+\frac{\mu^2}{\enth^2\belpot^2}\frac{M^2\left(\belpot^2-\belpa^2\right)^2}{\left(M^2+\belpot^2-1\right)^2}
\,, \\
D=\frac{1-M^2-\belpot^2}{M^2}u_s^2-\frac{\mu^2}{\belpot^2\enth^2}\frac{M^2 \left(\belpot^2-\belpa^2 \right)^2}{ \left(1-M^2-\belpot^2\right)^2} \nonumber\\
+\frac{M^2}{ \belpot^4 f^2 \enth^2}\left[ \frac{\belpot ^2-1}{\sin^2\left(\vartheta-\theta\right)}+M^2\right]
\,.
\end{eqnarray*}

The overall procedure of integration can be stated as follows. The differential Eqs.~(\ref{feq}, \ref{eqenth}, \ref{Meq}) together with the algebraic Bernoulli Eq.~(\ref{bernoulli}) and the initial conditions consist the system of equations that fully describe the flow. The initial conditions $f_0,M_0,h_0,\vartheta_0$ and the specific values of $\modind$, $\polind$ determining the evolution of the various quantities along the initial surface $\theta_0$, also provide the integrals $\massint,\belpot, \belpa^2 ,\mu,Q$ and complete the necessary set of parameters. The integration derives the evolution of the quantities along a specific poloidal streamline and then the similarity property $\bflux \propto \left(r / f \right)^\modind$ is used to extend this solution to the rest of the flow.

Two remarks are easily obtained by the straightforward inspection of Eq.~(\ref{Meq}). The rarefaction wave front occurs when the denominator vanishes, since these are the only points where the first derivatives might suffer a discontinuity. In order to give an intuitive interpretation, we write both $N,\, D$ in terms of the physical quantities:
\begin{eqnarray}
N=\frac{2p_{\rm total}}{\dens \enth c^2}\,, \quad p_{\rm total}=p+\frac{B^2-E^2}{8 \pi} \,, \quad \nonumber \\
\label{denom}
D=\frac{\left(\frac{\lrz v_\theta}{c}\right)^4-\left(\frac{\lrz v_\theta}{c}\right)^2\left(u_s^2+\frac{B^2-E^2}{4\pi\dens\enth c^2}\right)+u_s^2 \frac{B_\theta^2-E_r^2}{4\pi\dens\enth c^2}}{\left(\frac{\lrz v_\theta}{c}\right)^2} \,. \quad
\end{eqnarray}
The nature of the denominator vanishing becomes clear if we use both Eq.~(\ref{eqnprimedefB}) to rewrite the last term as $B_\theta^2-E_r^2=(1-\belpot^2)B_\theta^2$. The comparison with the dispersion relations for the magnetosonic disturbances, see Appendix C in \cite{Vlahakis_2003a}, reveals that the denominator vanishes when the $\hat\theta$ component of the flow proper velocity is equal with the comoving fast or slow magnetosonic phase velocity of a wave propagating along the $\hat\theta$ direction. These are the actual singular points of the steady-state flow \cite{Bogovalov_1994} forming the so-called modified fast/slow surface, or limiting characteristics, and it is already met in a number of approaches (see \cite{Tsigkanos_1996,Bogovalov_1997} and references therein).\footnote{Notice that a similar distinction for the Alfv\'enic point does not exist, since in the relevant dispersion relation the trigonometric/projecting terms cancel out.}

In the limit of vanishing poloidal magnetic field ($B_p \to 0$) a complication of interpretation nature enters. In such a case $\bflux$ becomes zero. Also $k$, $\belpot$, $M^2$ become infinite, but their ratio retains a finite value
\begin{eqnarray}
\frac{\belpot^2}{M^2}=\sigma\,, \quad \frac{M^2}{\massint^2}=\frac{4\pi \enth}{\dens}\,, \quad \massint \nbl\bflux = \nbl\vflux \,.
\end{eqnarray}
For that reason, the integration has to be performed for $\sigma$ rather than for $M^2$, as in \cite{Kostas_2013}. In general, one could use $\sigma$ instead of $M^2$ even when $B_p$ exists, by using the expressions:
\begin{eqnarray}
\label{eqnssigM}
\sigma=\frac{\belpot^2-\belpa^2}{M^2+\belpa^2-1}\,, \\
\frac{d\sigma}{d\theta}=-\frac{\sigma}{M^2+\belpa^2-1}\frac{dM^2}{d\theta}\,, \nonumber\\
\label{seq}
=-\frac{\sigma}{M^2+\belpa^2-1}\frac{\left(\modind-1\right) M^2}{\tan\left(\vartheta-\theta\right)}\frac{N}{D} \nonumber
\end{eqnarray}
(with the latter substituting Eq.~\ref{Meq}).

\section{Numerical Results}
\label{results}

Suppose a homogeneous magnetized plasma having magnetic field $B_{p0}\hat z + B_{y0} \hat y$ and bulk velocity $v_{p0}\hat z + v_{y0} \hat y$ fills the space $z<0$, $x<0$, supported by some external pressure on the plane $x=0$, $z<0$, see Fig.~\ref{geom}. Our goal is to explore how a sudden pressure drop at the origin $x=z=0$ modifies the flow characteristics in the region $z>0$, through the rarefaction wave that propagates as a weak discontinuity inside the body of the flow.

Since we require the initial flow to be homogeneous we fix the parameter $\modind=1$. We chose three different set of initial configurations for cold super-fast magnetosonic flows given in Table~\ref{arrayinitial}. In relation to the strength of the poloidal magnetic field we include cases in which: (i) the poloidal magnetic field component is negligible ($B_p \ll |B_y|$, LP model), (ii) the poloidal magnetic field is mildly smaller than the transverse one ($B_p < |B_y|$, MP model), (iii) both components are of similar magnitude ($B_p \sim |B_y|$, EP model). Our attempt to increase further the strength of the poloidal magnetic field is restricted by the condition of staying in the super-fast magnetosonic regime.
We also include a fourth model (TD) in which the thermal energy is nonnegligible.
The implications of the initial transverse velocity are studied in the remaining two models (LP01, LP03). These are the same as the cold flow model (LP) except their initial transverse velocities ($v_{y0} = 0.1, \, 0.3$ respectively).

The results of the integration show that both the poloidal poloidal magnetic field and the initial transverse velocity affect the spatial scale of acceleration as also the rarefaction wave front inclination. Finally, and in order to compare our results with the ones obtained by numerical simulations, we included models (HDB), (MHDA), (MHDB) that stand for the corresponding scenarios simulated in \cite{Mizuno_2008}. The initial parameters for these models are shown in Table~\ref{arrayinitialMiz}.

\begin{table}
\begin{tabular}{l|cccc} 
    Model & $\sigma_0$ & $-(B_y/B_p)_0$ & $v_{y0}$  &$M_0$ \\ 
    \hline\hline
    low poloidal (LP)  &  $10$ & $40000$ & $0.0005$ & $12000$\\ 
    mild poloidal (MP) &  $10$ & $158$ & $0.0005$ &  $50$\\ 
    equal poloidal (EP) &  $10$ & $3.5$ & $0.0005$  & $1.10$\\ 
    thermal driven (TD) &  $0.1$ & $0.6$ & $0.01$  &$2.0$ \\ 
    low poloidal01 (LP01) & $10$ & $40000$ & $0.1$ & $12000$\\ 
    low poloidal03 (LP03) & $10$ & $40000$ & $0.3$ & $11300$\\ 
    \hline
\end{tabular}
\caption{The initial conditions of our models were set at $\theta_0=-\pi/2$. All models represent cold flows $\enth=1$, except (TD) which is actually a thermally dominated one with $\enth_0=10$ and $\polind=4/3$, share the same total energy flux $\mu=1100$, the same initial Lorentz factor $\lrz_0=100$, are homogeneous $\modind = 1$, and the poloidal streamlines (and field lines since $\bm B_p \parallel \bm v_p$) are initially parallel to the $z$-axis ($\vartheta_0=0$).
\label{arrayinitial}}
\end{table}

The initial conditions for all models were specified at $\theta_0=-\pi /2$ (i.e. at $z=0$, $x<0$). The main criterion over which they were selected was the total energetic context
\begin{equation}
    \mu=\lrz\enth\left(1+\sigma\right) \,.
\end{equation}
All models shown in Table~\ref{arrayinitial} share the same value of $\mu = 1100$. The models of the cited simulations (Table~\ref{arrayinitialMiz}) do not share this same value; the corresponding fluxes are shown in the relevant column of that Table. The thermally driven models (TD and HDB) have very high enthalpy ($\enth_0=10,\, 21.6$) and thus a polytropic index of $\polind=4/3$ was chosen.

\begin{figure*}
        \includegraphics[width=0.3\textwidth,angle=0]{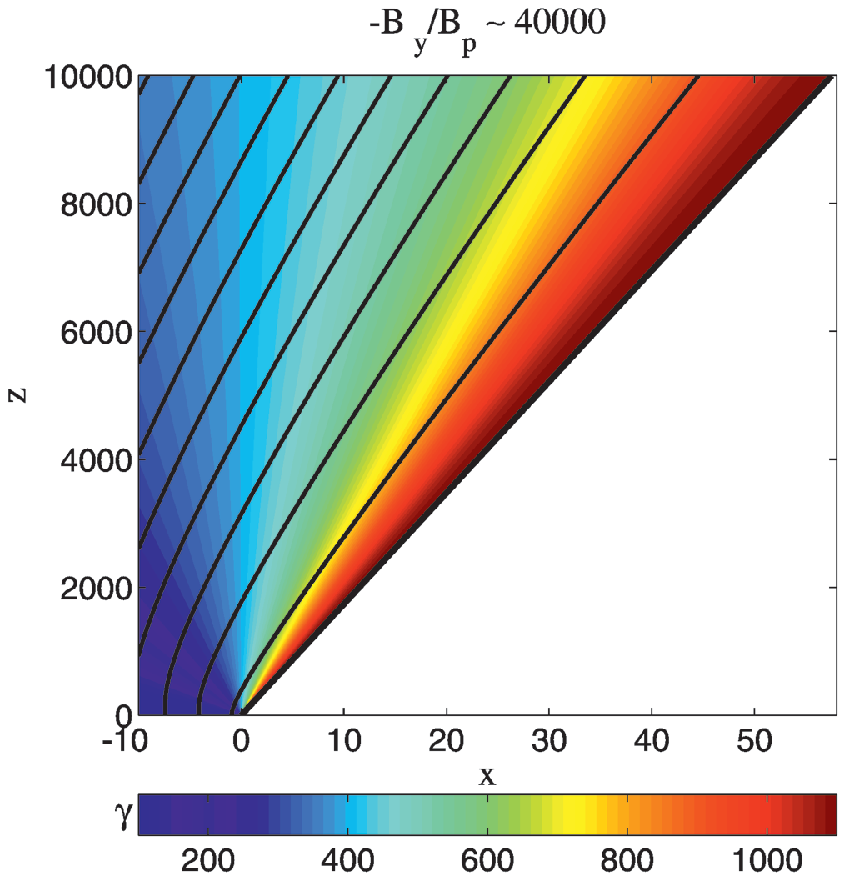}
        \includegraphics[width=0.3\textwidth,angle=0]{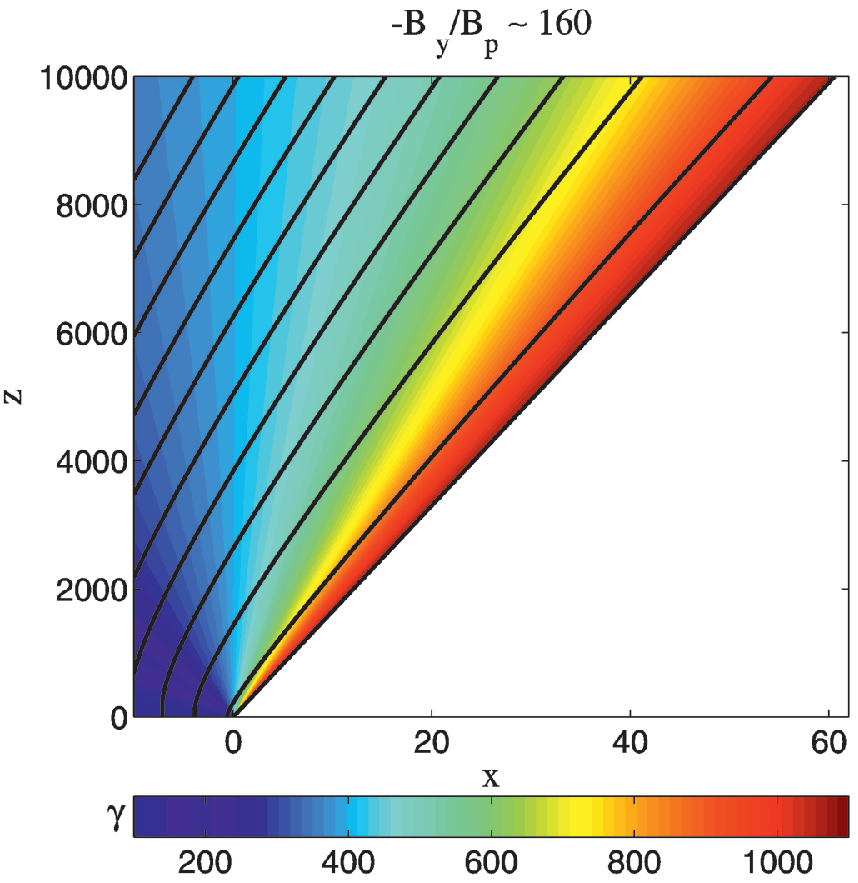}
        \includegraphics[width=0.3\textwidth,angle=0]{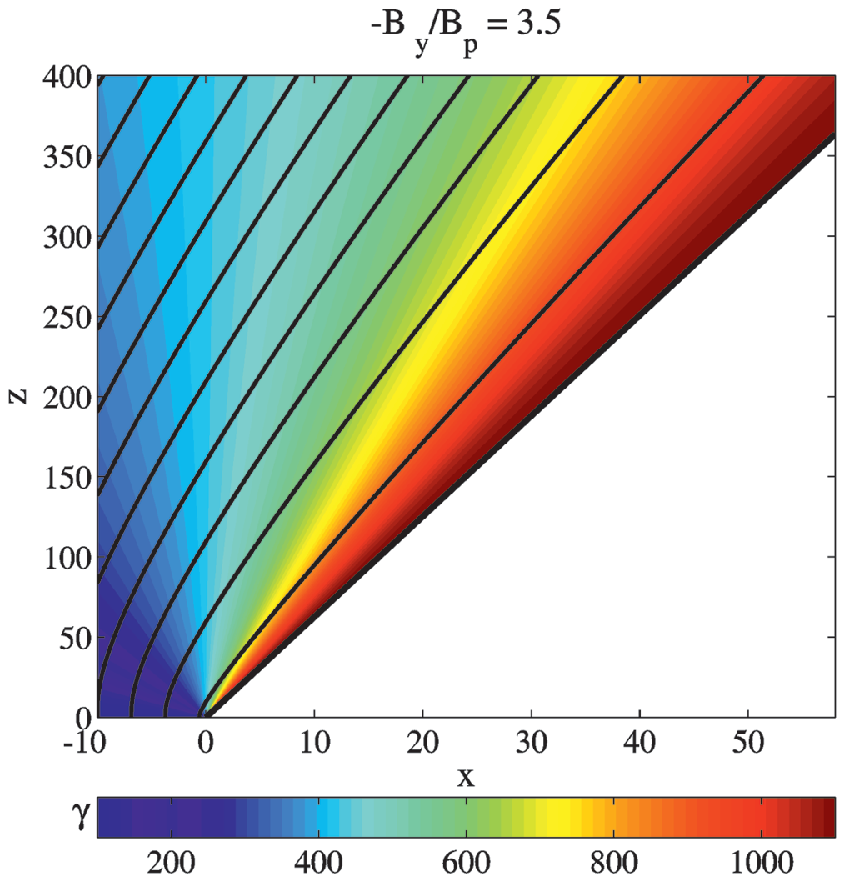}
        \\
        \includegraphics[width=0.3\textwidth,angle=0]{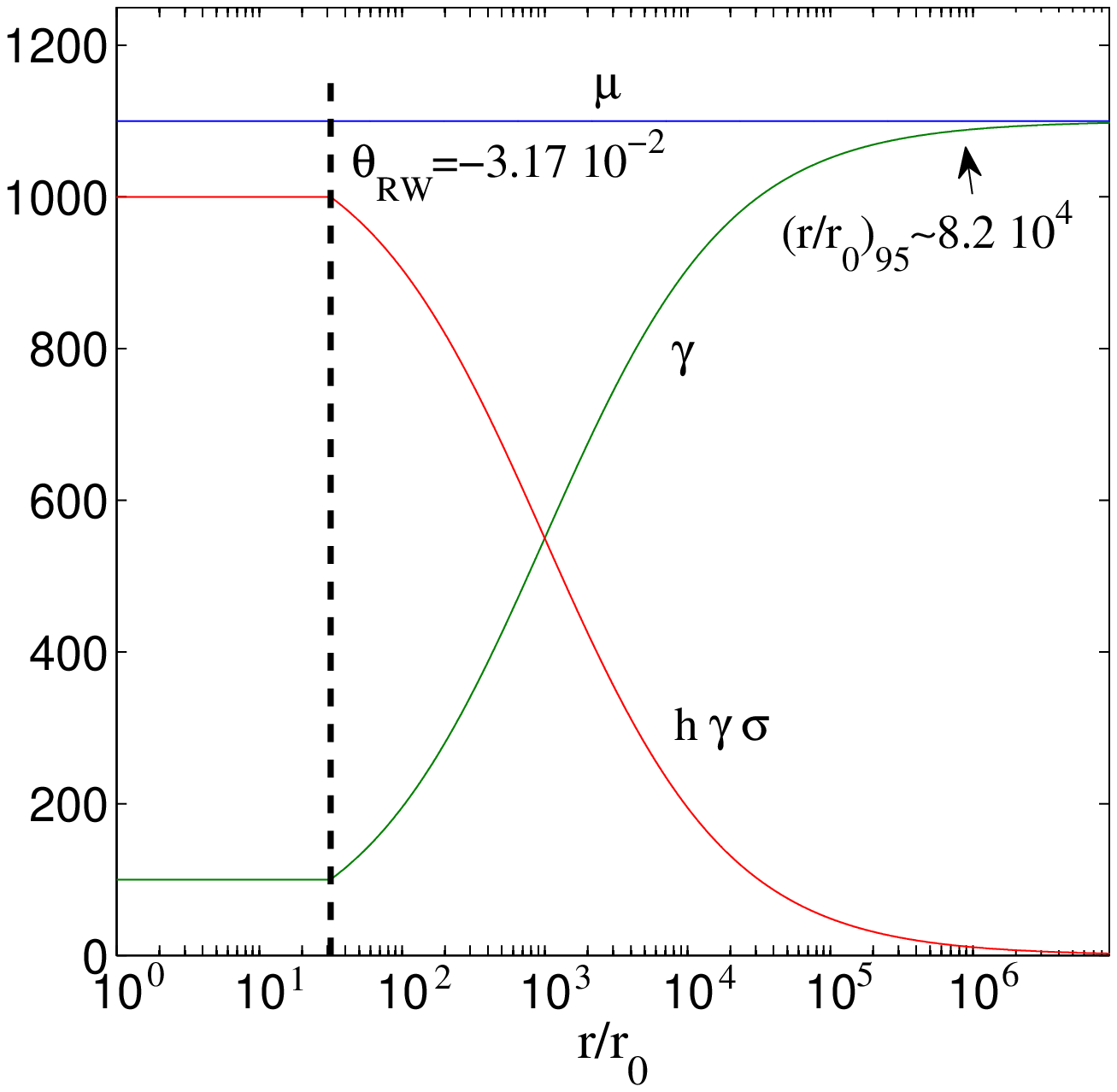}
        \includegraphics[width=0.3\textwidth,angle=0]{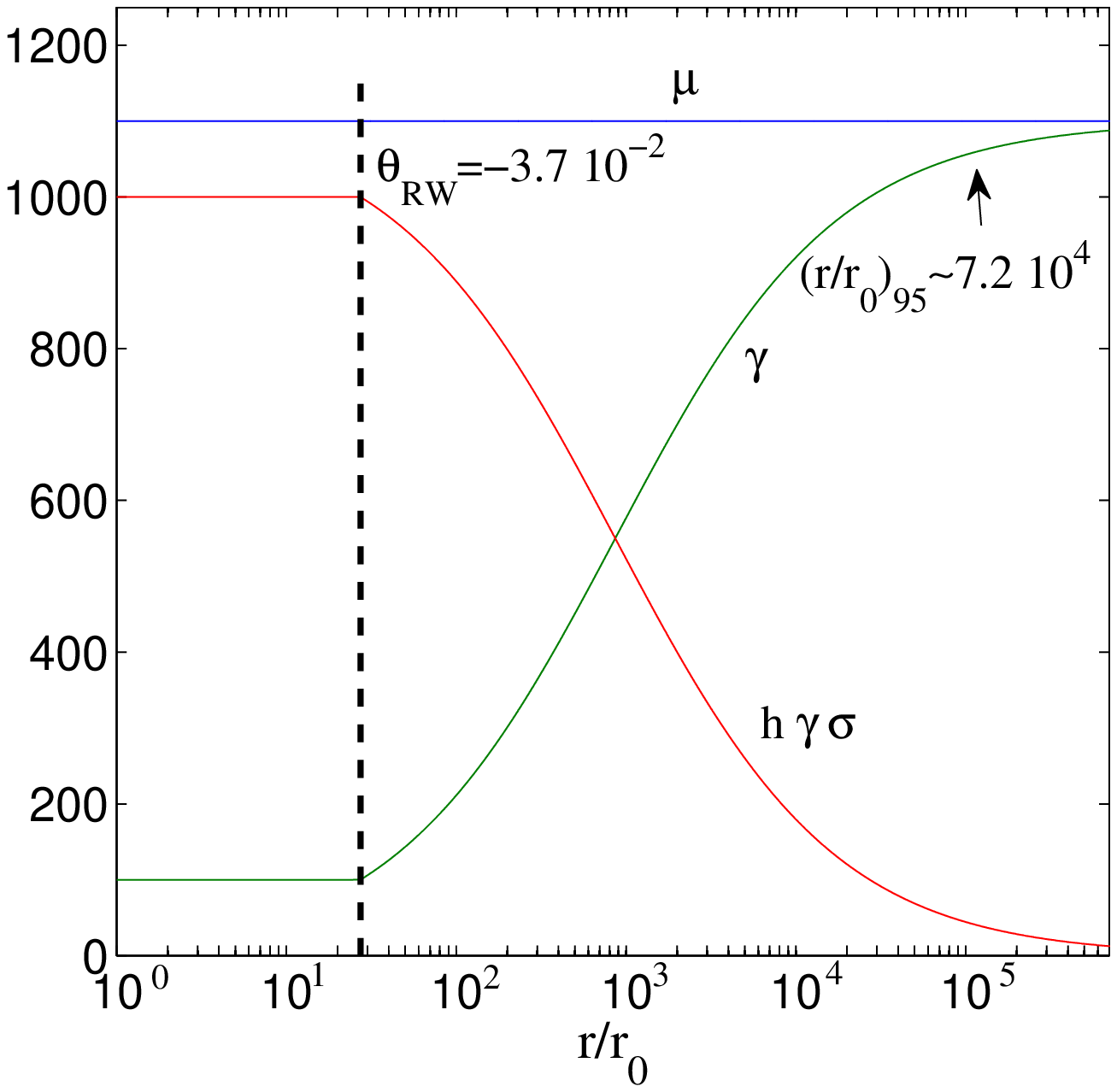}
        \includegraphics[width=0.3\textwidth,angle=0]{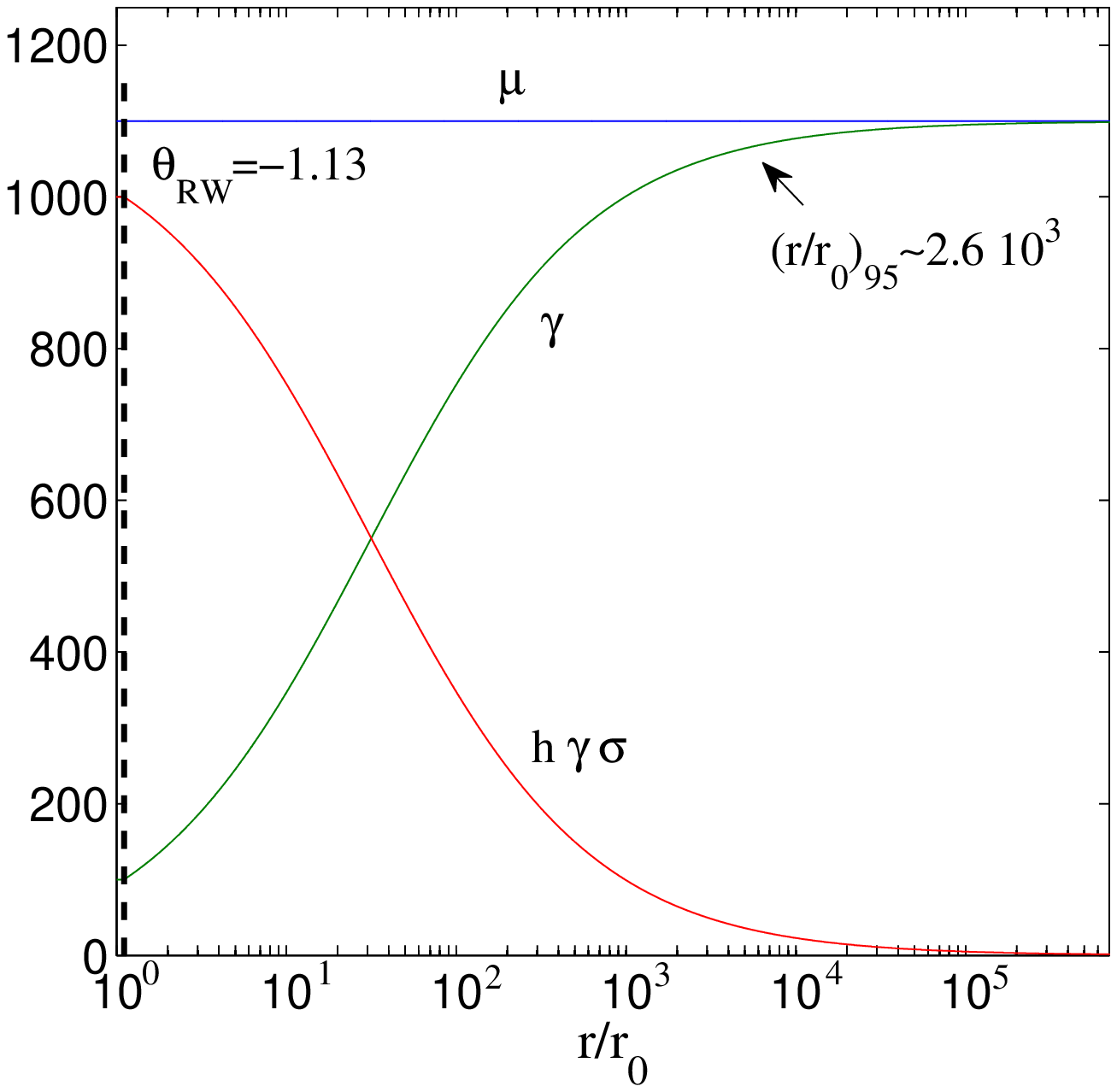}
        \\
        \includegraphics[width=0.3\textwidth,angle=0]{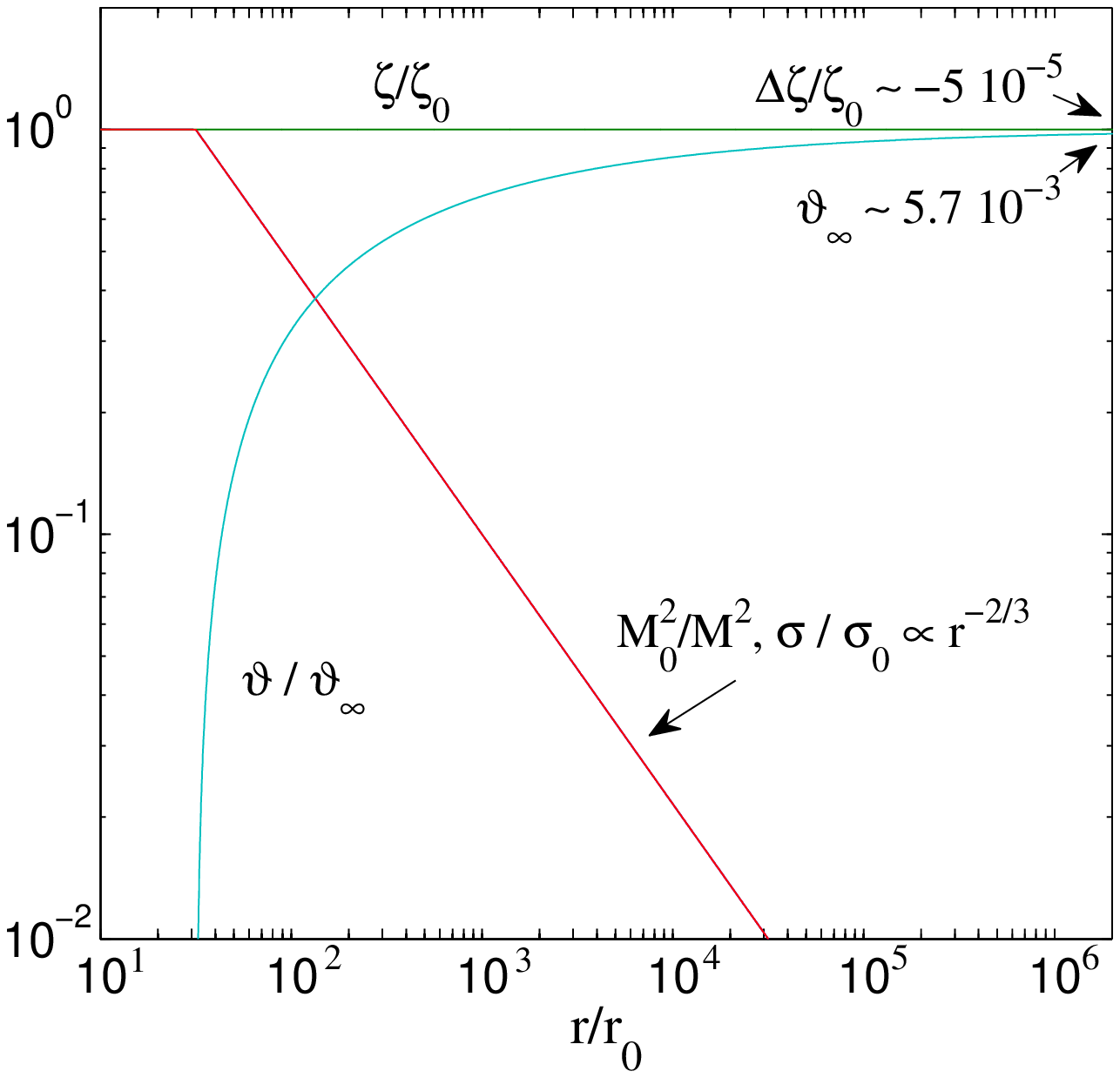}
        \includegraphics[width=0.3\textwidth,angle=0]{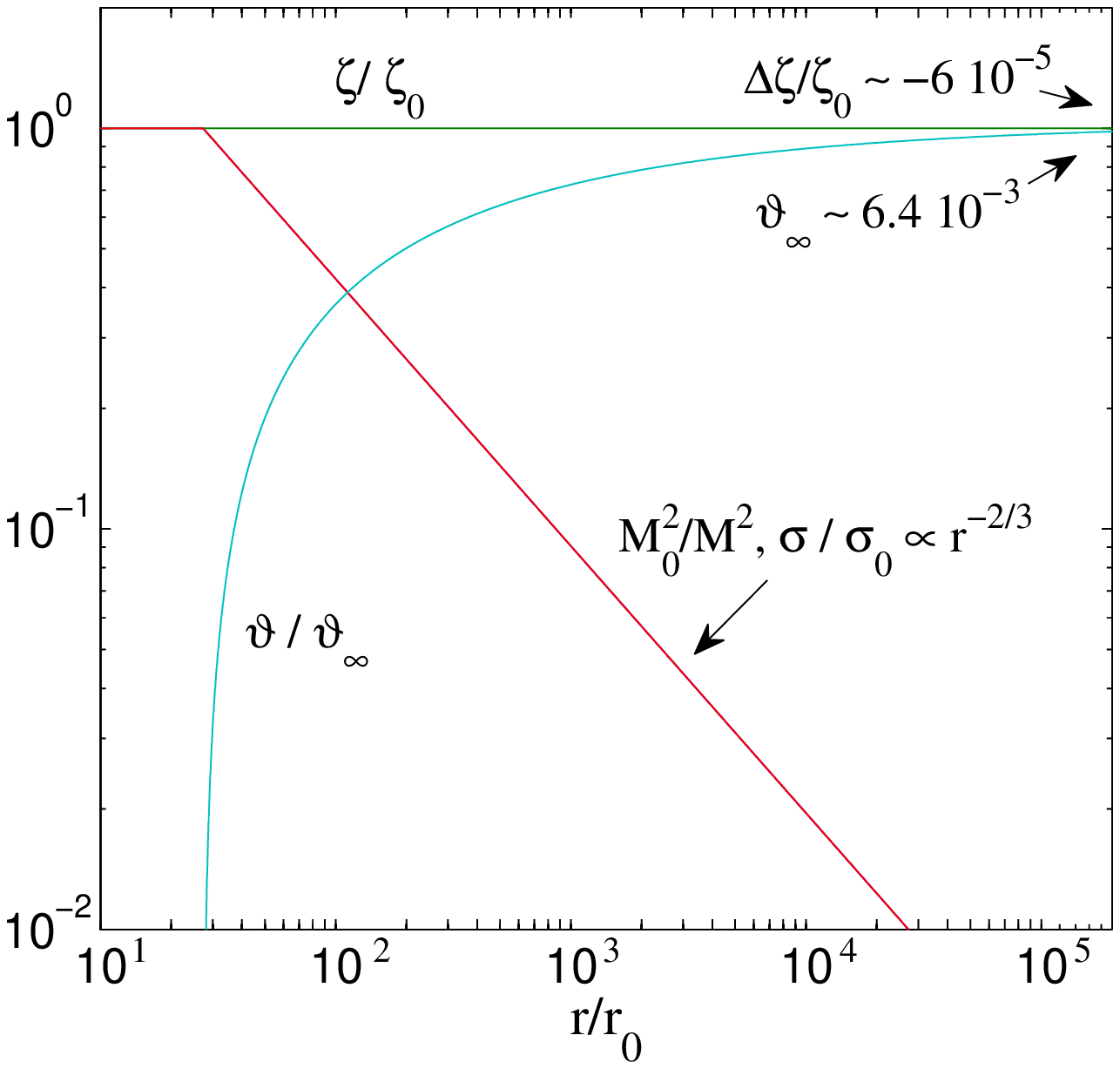}
        \includegraphics[width=0.3\textwidth,angle=0]{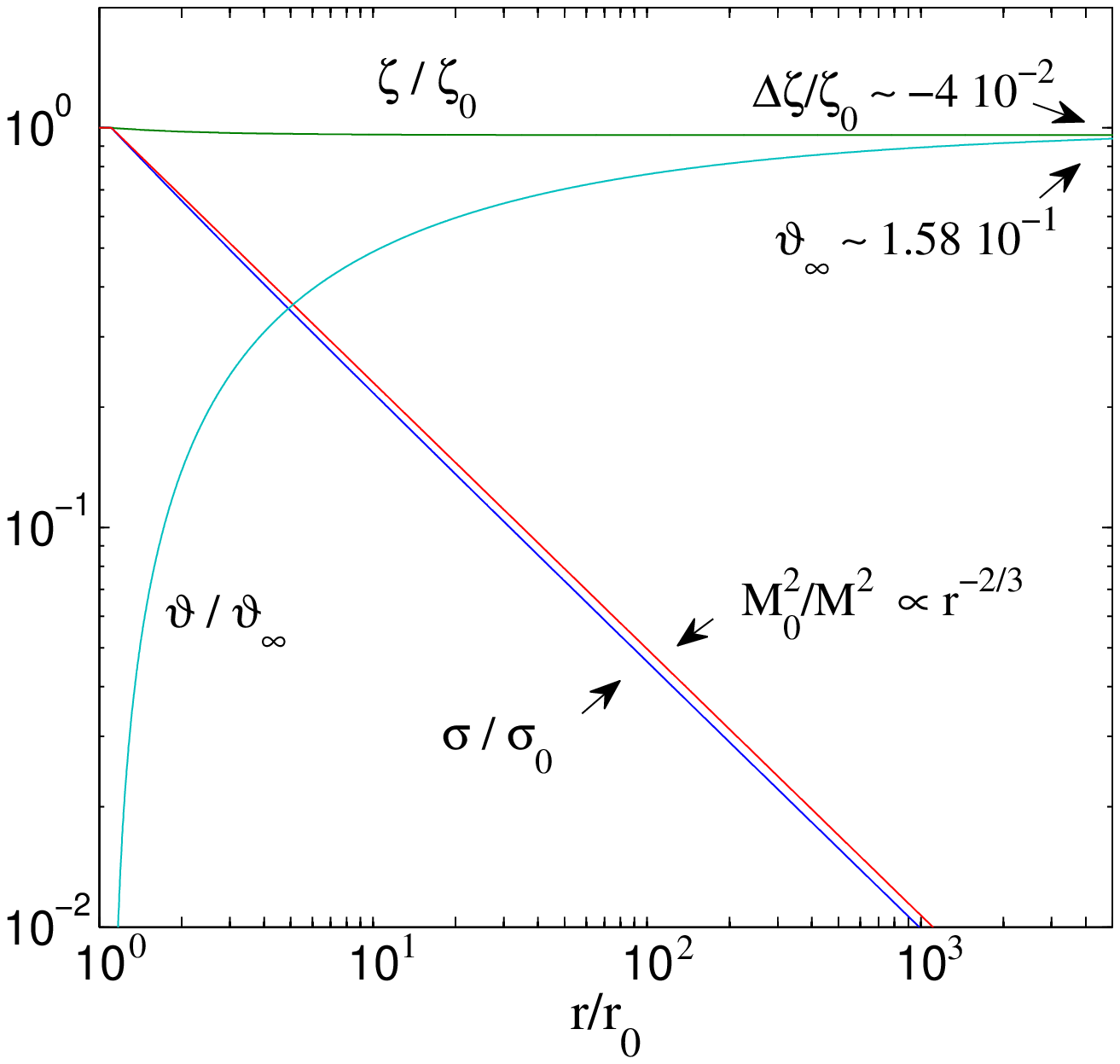}
        \\
        \includegraphics[width=0.3\textwidth,angle=0]{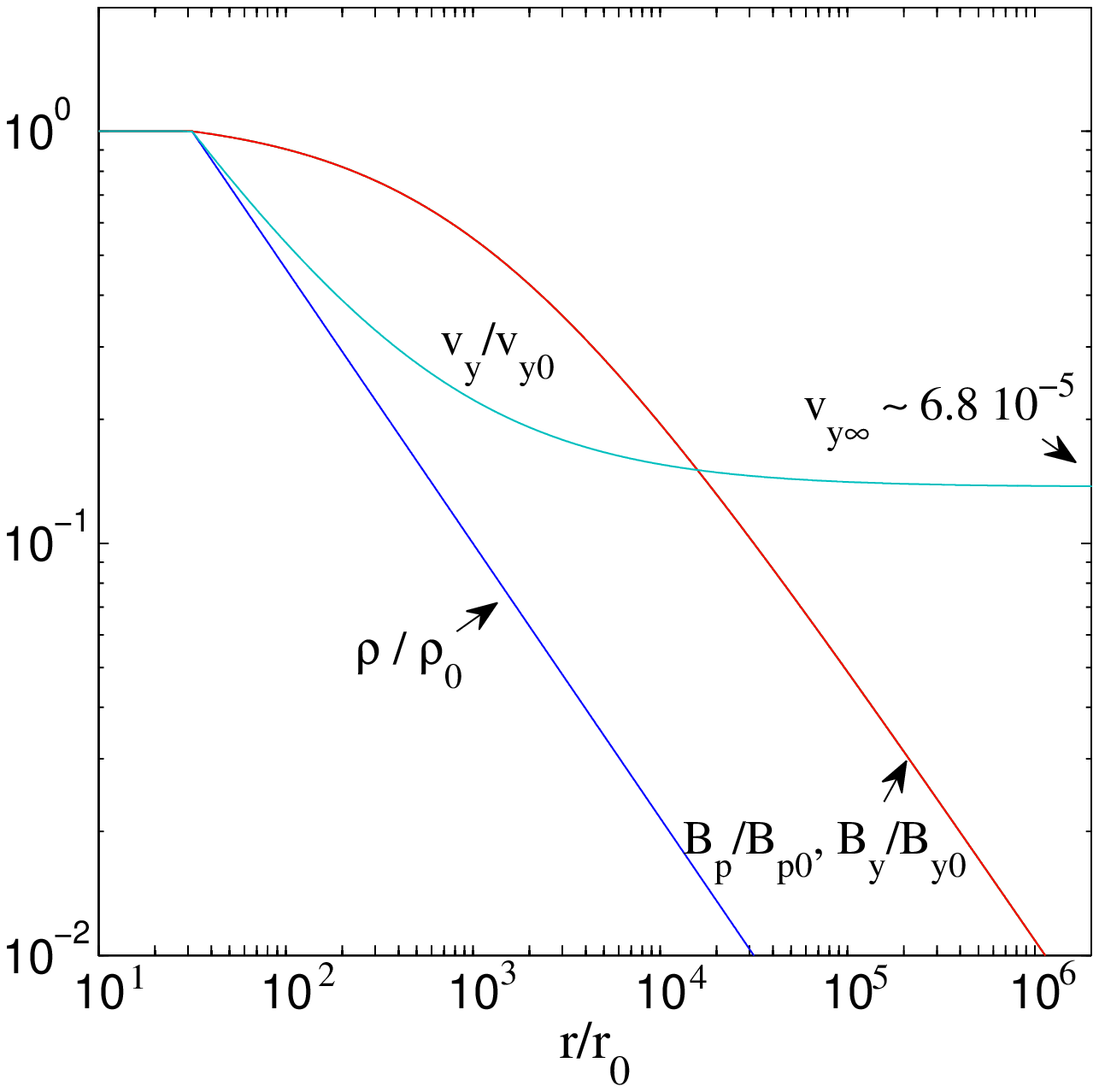}
        \includegraphics[width=0.3\textwidth,angle=0]{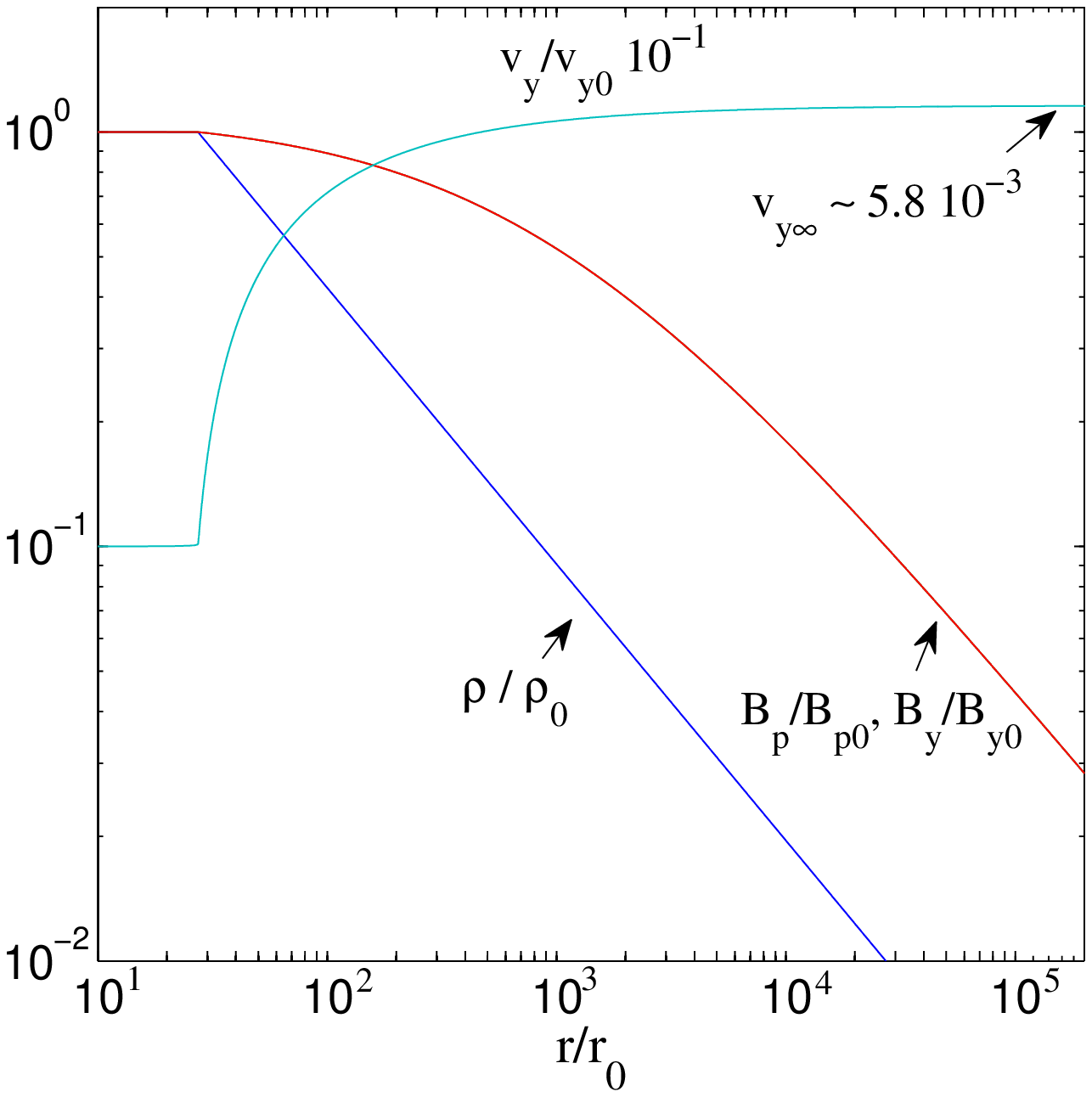}
        \includegraphics[width=0.3\textwidth,angle=0]{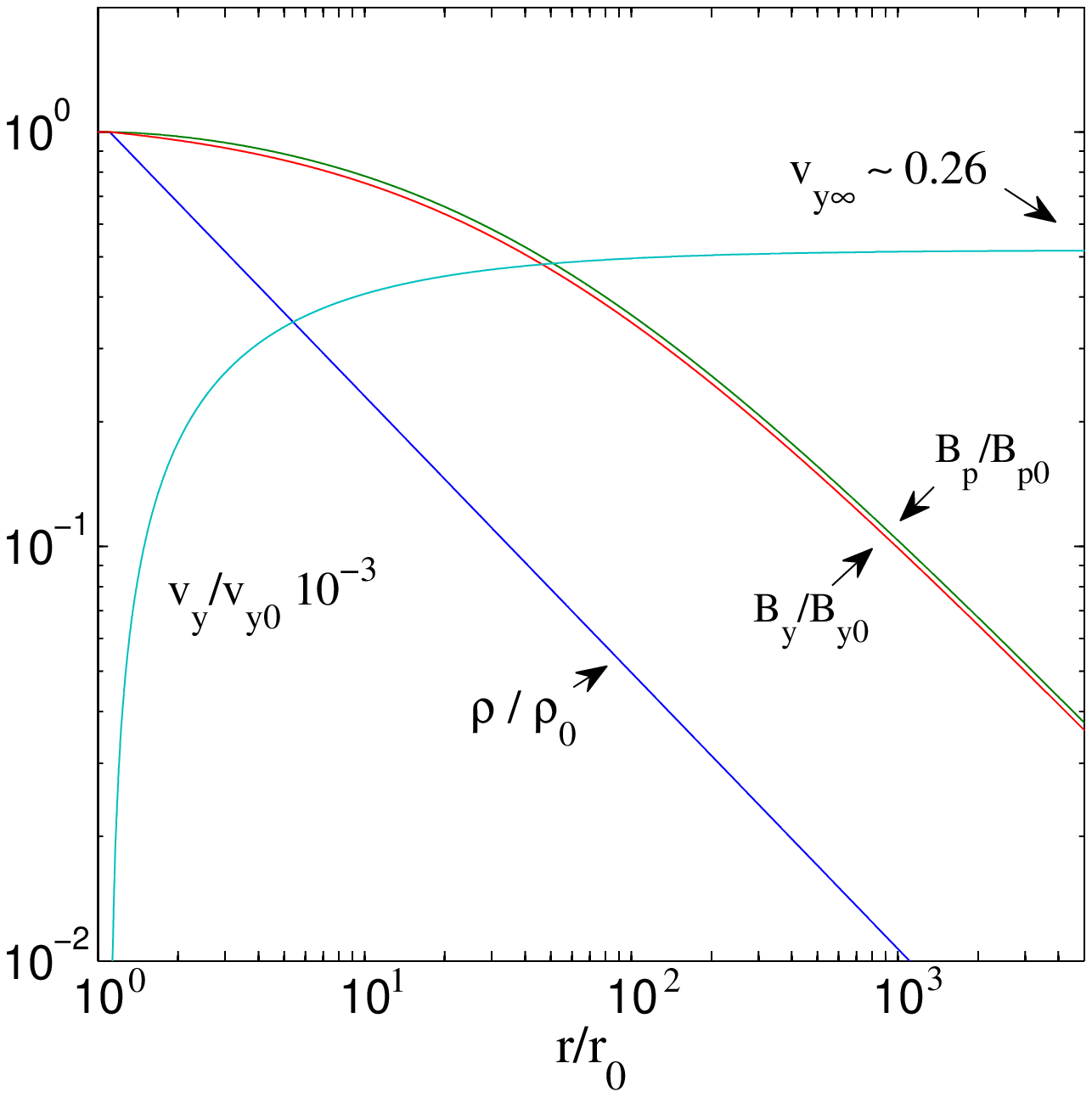}
    \caption{The results of the three cold models (LP, MP, EP). \textbf{First row.} The physical shape of the flow, some specific streamlines, and the spatial distribution of the Lorentz factor (color). \textbf{Second row.} The energetics along a specific line, where $\lrz$ (green) is the Lorentz factor, $\mu$ (blue) is the total energy flux and $\enth \lrz \sigma$ (red) the Poynting flux, both per rest mass energy flux. The dashed lines stand for the rarefaction wave front, and $(r/r_0)_{95}$ the point where the $\lrz$ reaches the $95\%$ of $\mu$. \textbf{Third row.} The evolution of the integrable functions normalized to the values mentioned in the diagrams. The inverse square of the Alfv\'enic Mach number (red) and the magnetization parameter (blue) follow the same power law in the first two models, while they slightly differ in the last one. The inclination of the poloidal streamlines (cyan) and its final asymptotic value ($\vartheta_\infty$) appears, the relative strength of the magnetic field components (green) and its asymptotic relative difference ($\Delta\ratb=\ratb_\infty-\ratb_0$). \textbf{Bottom row.} The evolution of the physical quantities normalized to their initial values. The density ratio ($\dens / \dens_0$, blue). The evolution of the magnetic field components ($B_p/B_{p0}$ red, $B_y/B_{y0}$ green) coincides in the first two diagrams and differing slightly at the last one. The normalized transverse velocity ($v_y/v_{y0}$, cyan) and its final asymptotic value also appears; notice its negligible value at the first two models in contrast to the last one.}
    \label{figcoldmods}
\end{figure*}

The results for the magnetic dominated models (LP, MP, EP) appear in Fig.~\ref{figcoldmods}
The first row of diagrams shows the physical shape of the flow, some streamlines, and the spatial distribution of the Lorentz factor. One must have in mind that if a line attains a specific value of the Lorentz factor at some point, self-similarity will finally ascribe this efficiency and to the rest lines starting from the lines close to the corner to the most exterior ones. So the suitable measure for the efficiency achieved is not the efficiency itself, but the relevant energetic evolution along a streamline as a function of the angle $\theta$ or similarly of the relative distance $r/r_0=f/f_0$, where $r_0$ the initial radial distance that the line originates from the base of the flow ($\theta_0=-\pi/2$) and $f_0$ the value of $f$ at that point.

The energetics are shown in the second row, where we draw the energy fluxes per mass energy flux in the laboratory frame. The Lorentz factor for a cold flow is equal to the inertial energy flux (rest mass energy plus bulk kinetic), while the thermal energy $\left( \enth-1\right) \lrz$ is absent. During the rarefaction evolution the Poynting flux is converted efficiently to kinetic, reaching soon to its maximum possible value ($\lrz_{max} = \mu$); $\left(r/r_0\right)_{95}$ the point where $\lrz$ attains $95\%$ to its maximum value. When $B_p$ becomes comparable to $|B_y|$ it has significant impact both on the rarefaction wave front and the spatial scale of acceleration in the last one. The analytical results obtained in Appendix~\ref{appA} interpret exactly this behavior, a summary of which is shown in Fig.~\ref{thapp}. The calculated rarefaction front corresponds to the dashed lines in the second row diagrams.

An important conclusion of our study appears in the third panel exhibiting the evolution of the integrable quantities along the streamline. The inverse of the Alfv\'enic Mach number and the magnetization parameter follow a power law decrease ($M^{-2}, \sigma \propto r^{-2/3}$), and only small deviations are noticed in the EP model. The scale of the Alfv\'enic Mach number is associated to the planar geometry, and is further discussed in the next section as also in Appendix~\ref{appA} where a formal derivation of the scaling law is given. The extend of the rarefied region, equals to the so-called Prandtl-Meyer angle which is defined as the angle between the initial and the final orientation of the flow ($\theta_{PM} = \vartheta_\infty \sim \theta_\infty$). Its monotonic increase with decreasing magnetic field component ratio $\ratb_0$, with $\ratb\equiv -B_y / B_p$, is also provided in Appendix~\ref{appA}, Eq.~(\ref{eqnassympPM}), and demonstrated in Fig.~\ref{thapp} for low initial transverse velocities. As for the relative magnetic field component strength included in the diagrams we notice that the ratio doesn't alter much, although in the last case  a small difference exists that wouldn't be significant if it didn't had serious implications on the derivation of the expression providing $\theta_{PM}$ angle (see in Fig.~\ref{thapp} how sensitive is the value of $\theta_{PM}$ as a function of $\ratb_0$, for not too high values of $\ratb_0$).

The physical quantities normalized to their initial values appear in the last row. The decrease observed in all except the transverse velocity, is intuitively expected due to the rarefaction process and the relevant conversion of the Poynting energy. The density decrease follows also the $-2/3$ power law, as Eq.~(\ref{eqnsdens}) suggests ($\dens \propto M^{-2}$). Similar behavior follow the magnetic field components, where only small deviations at the low $\ratb_0$ cases exist. Some special attention should be given in the transverse velocity evolution that shows a peculiar pattern either of increase (LP), or decrease (MP, EP), explained in the next section.

\begin{figure*}
        \includegraphics[width=0.3\textwidth,angle=0]{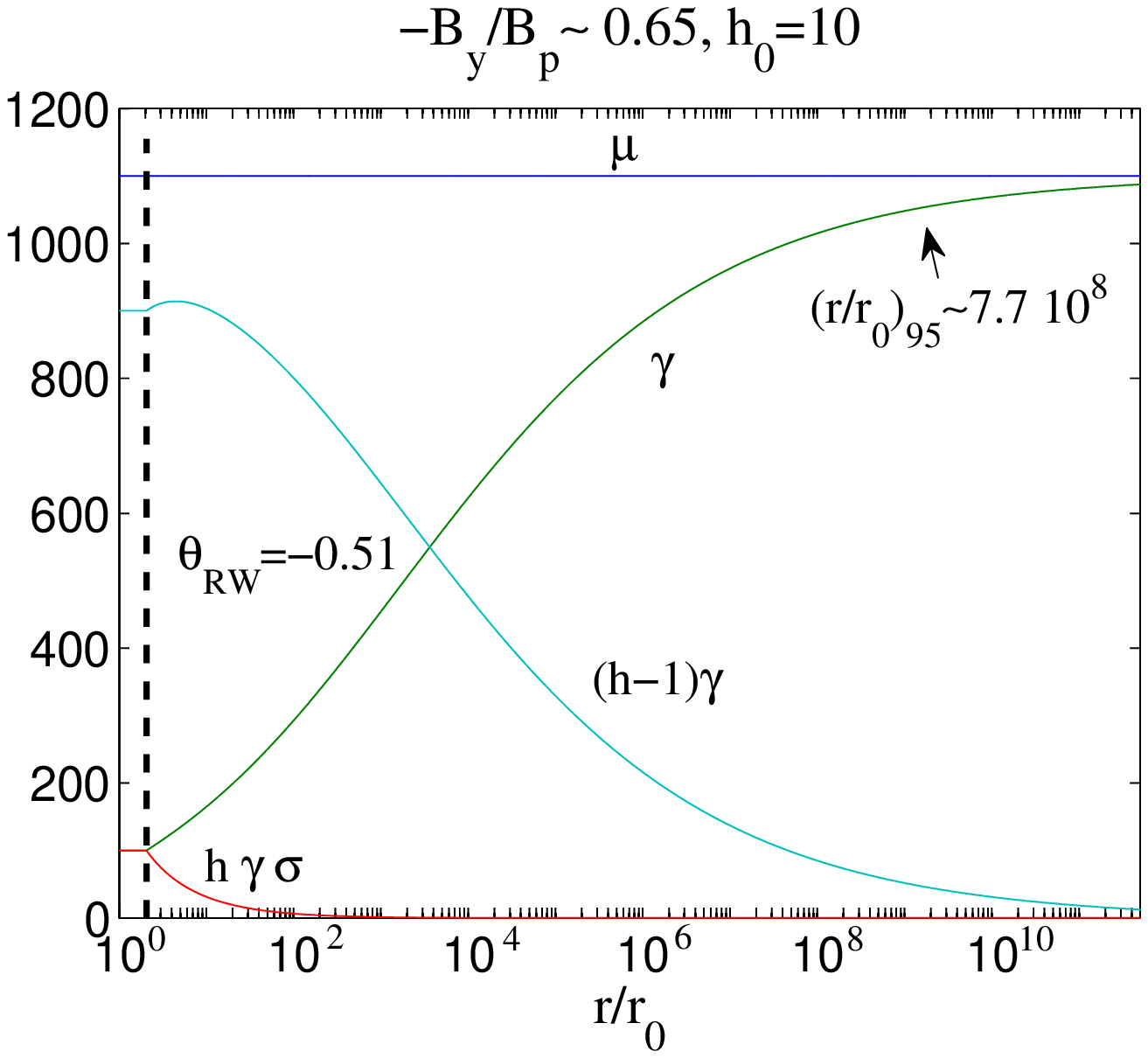}
        \includegraphics[width=0.3\textwidth,angle=0]{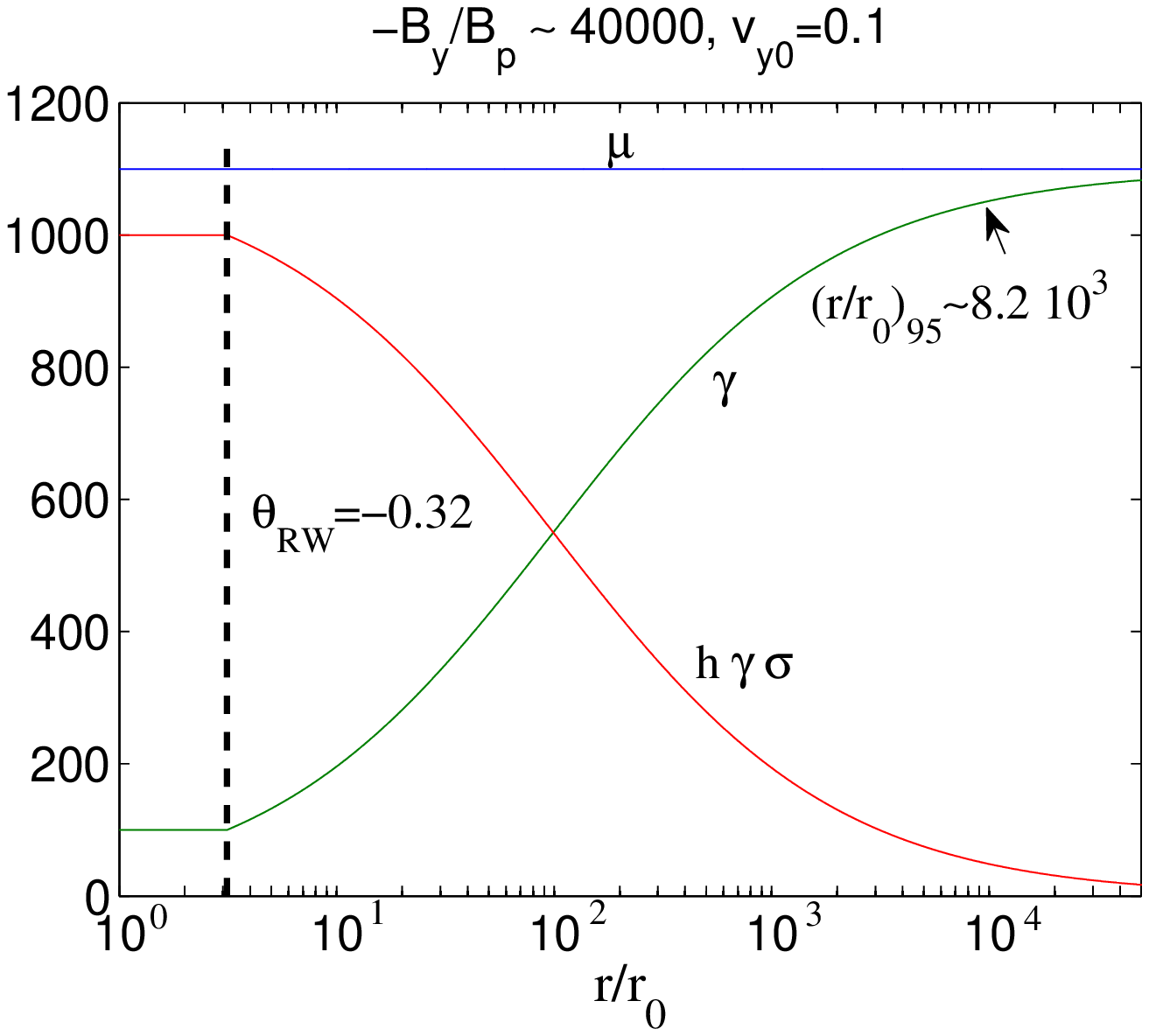}
        \includegraphics[width=0.3\textwidth,angle=0]{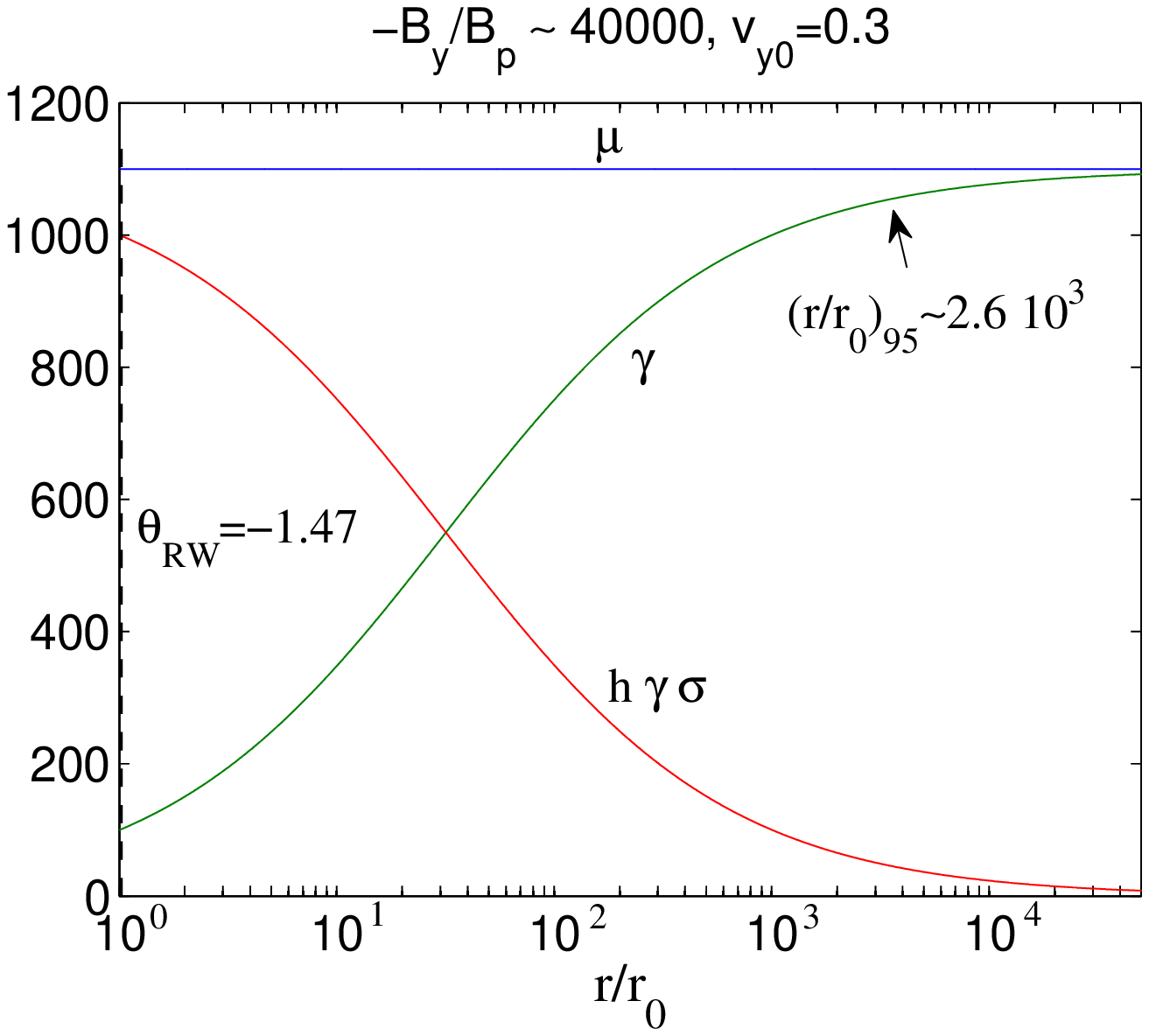}
        \\
        \includegraphics[width=0.3\textwidth,angle=0]{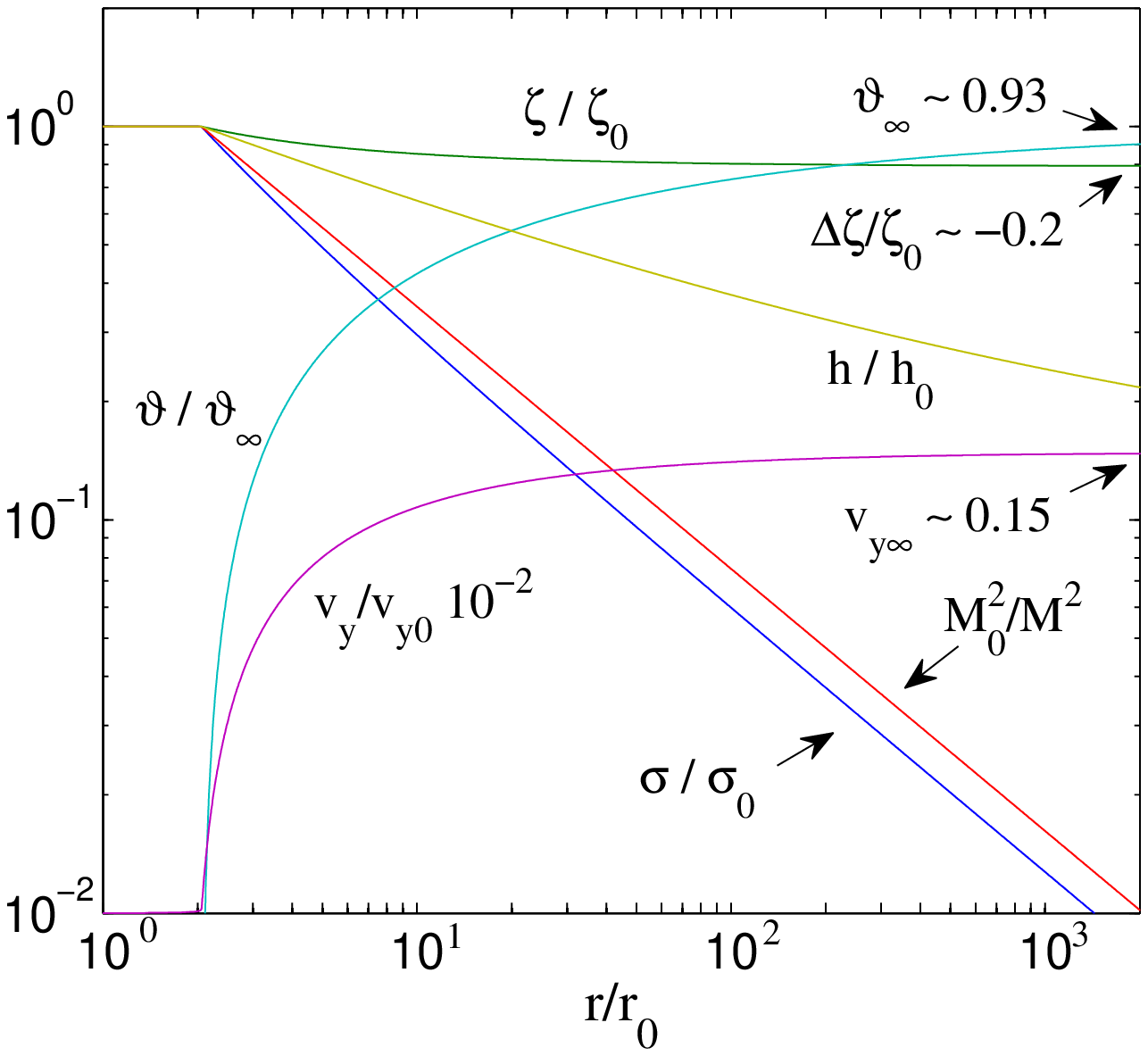}
        \includegraphics[width=0.3\textwidth,angle=0]{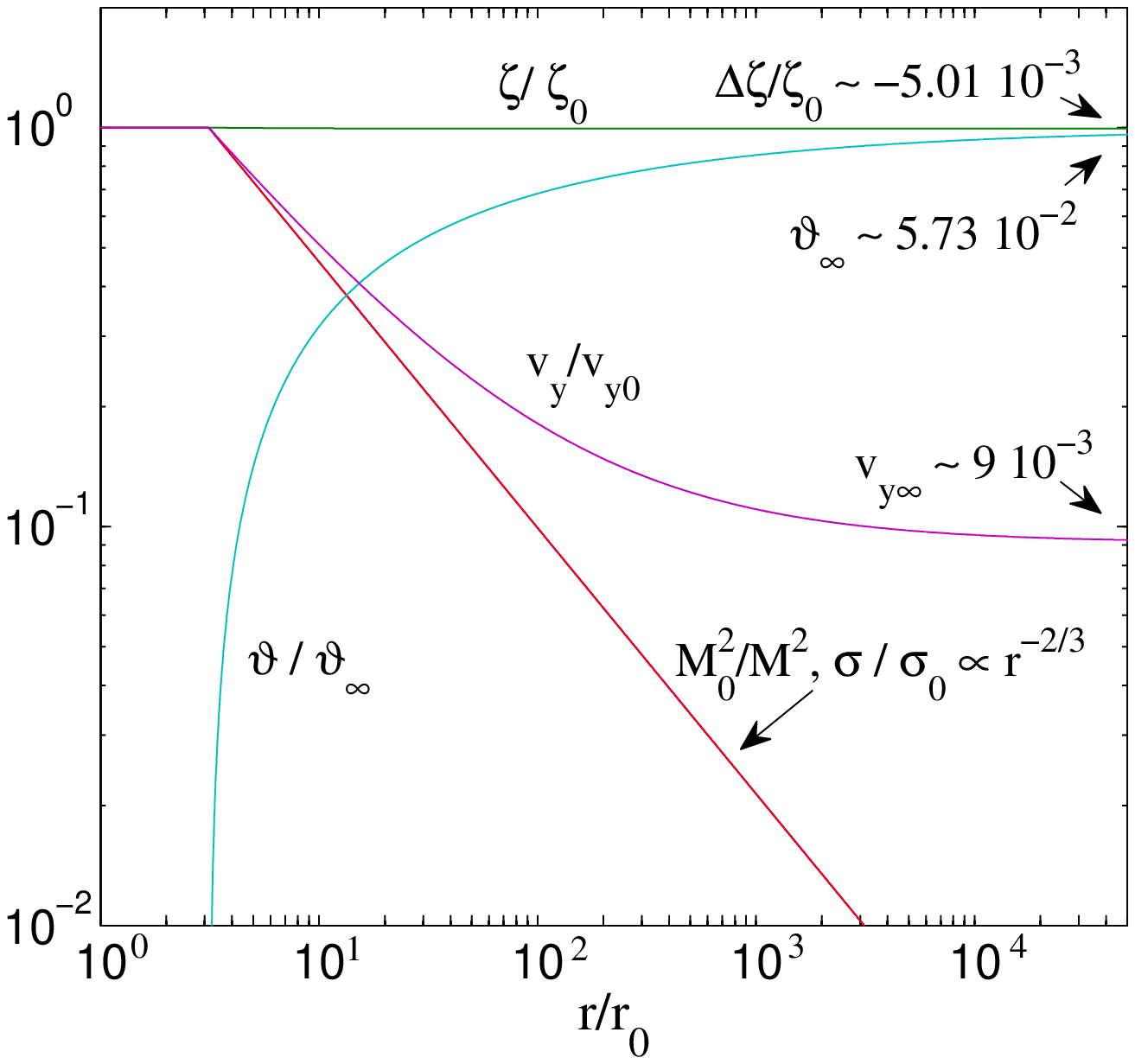}
        \includegraphics[width=0.3\textwidth,angle=0]{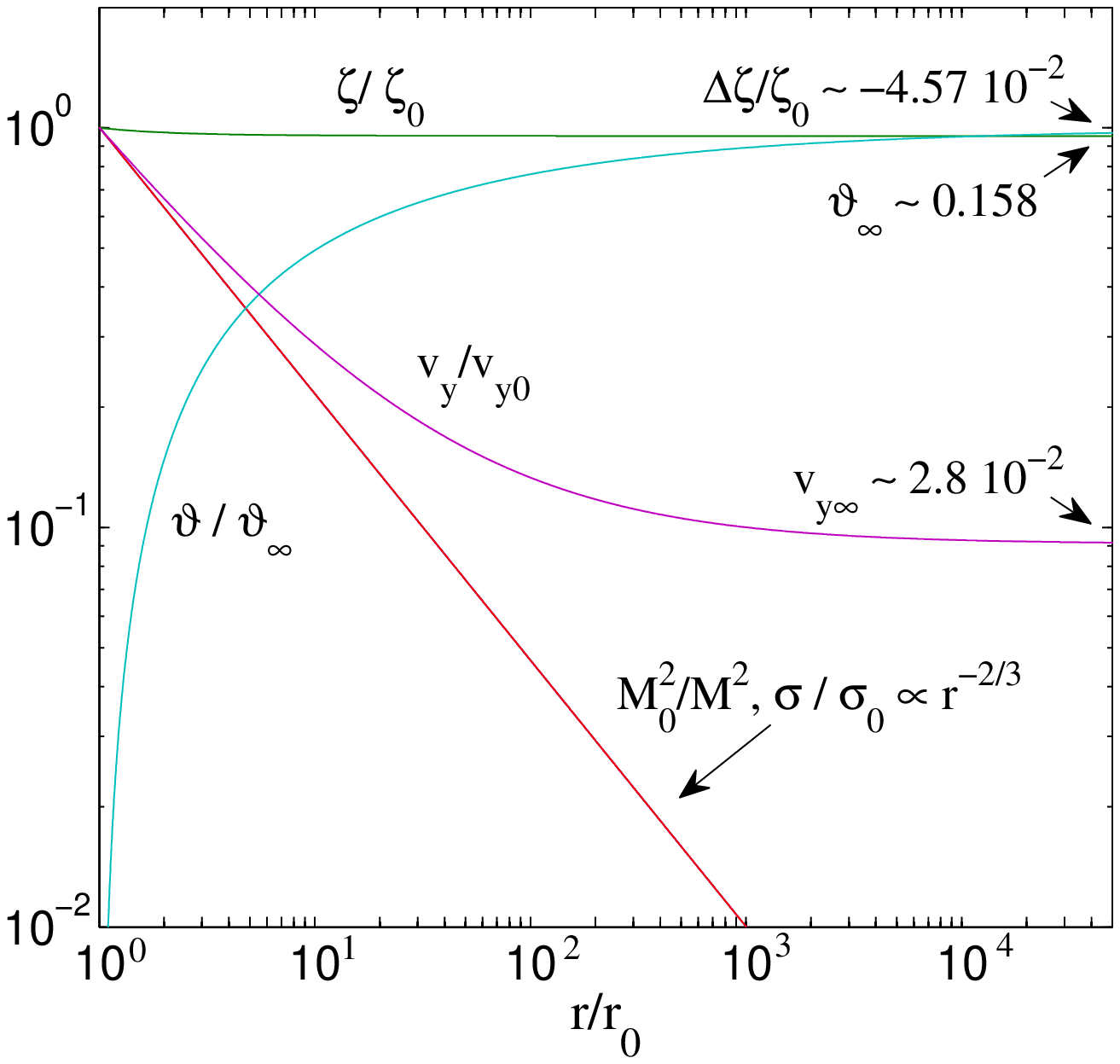}
    \caption{The results for the TD, LP01, LP03 models. \textbf{First row.} The total energy ($\mu$ blue), the Lorentz factor ($\lrz$ green), the Poynting flux ($\enth \lrz \sigma$ red) and the thermal one ($\left(\enth-1\right)\lrz$ cyan); the two last models corresponds to cold flows and thus the thermal energy does not appear. The dashed line corresponds to the calculated distance where rarefaction wave occurs. \textbf{Second row.} The evolution of some important quantities normalized in the values shown at the diagrams. The inverse square of the Alfv\'enic Mach number (red), the magnetization parameter (blue), the inclination of the of the poloidal filed lines (cyan), and the relative transverse velocity (magenta); the asymptotic values of the last two also appear. The ratio of the magnetic field components (green), and its relative asymptotic difference ($\Delta \ratb=\ratb_{\infty}-\ratb_0$). Notice that in the (TD) model, we included the specific enthalpy evolution (yellow) to exhibit its effects on the thermal driven rarefaction.}
\label{figrestmods}
\end{figure*}

The results of the last three models (TD, LP01, LP03) appear in Fig.~\ref{figrestmods}. The main characteristics of the mixed type scenario is the much lengthier spatial scales of acceleration, the  latter appearance of the wave front, and the extension of the rarefied region; compare for example with (EP) model. A small bump observed in the thermal energy curve  $(\enth-1)\lrz$ is due to the magnetic acceleration of the flow. That acceleration yields an increases of the inertial of the thermal energy rather than of the thermal energy context, see that $\enth$ monotonically decreases in the bottom row. This first phase of acceleration occurs in expense of the Poynting energy and for that reason the Alfv\'enic Mach number follows the power law scaling mentioned before. That behavior breaks when the thermal energy becomes the leading one, but this region falls out of the diagram.

Models (LP01, LP03) are dedicated to the implications of significant initial transverse velocity in contreast to the (LP) model. As seen in Fig.~\ref{figrestmods} the effects on the rarefaction wave inclination and on the maximum extension of the rarefied area are important, while $M^{-2}$ and $\sigma$ still follow the $-2/3$ power law. Besides its high initial value, the transverse velocity finally declines to small values asymptotically.

\begin{table}
\begin{tabular}{l|cccc}
    Model & $\sigma_0$ & $B_{p0}$ & $-B_{y0}$ & $\mu$ \\
    \hline\hline
    poloidal field (MHDA)  & $0$ & $21.27$ & $0$ & $2.83\times 10^{6}$\\
    transverse field (MHDB) & $100$ & $0$ & $149$ & $5.38\times 10^{5}$\\
    hydrodynamic flow (HDB) & $0$ & $0$ & $0$ & $4.11\times 10^{4}$\\
    \hline
\end{tabular}
\caption{All models share the same initial velocity along the $z$-axis, with $\lrz_0=7.089$ ($\vartheta_0=0$, $v_{z0}=0.990$, $v_{y0}=0$). The thermal energy is also common, with $\enth_0=4\times 10^5$ ($\dens_0=10^{-4}$, $p_0=10$, $\polind=4/3$).
\label{arrayinitialMiz}}
\end{table}

\begin{figure*}
    \includegraphics[width=0.25\textwidth,angle=0]{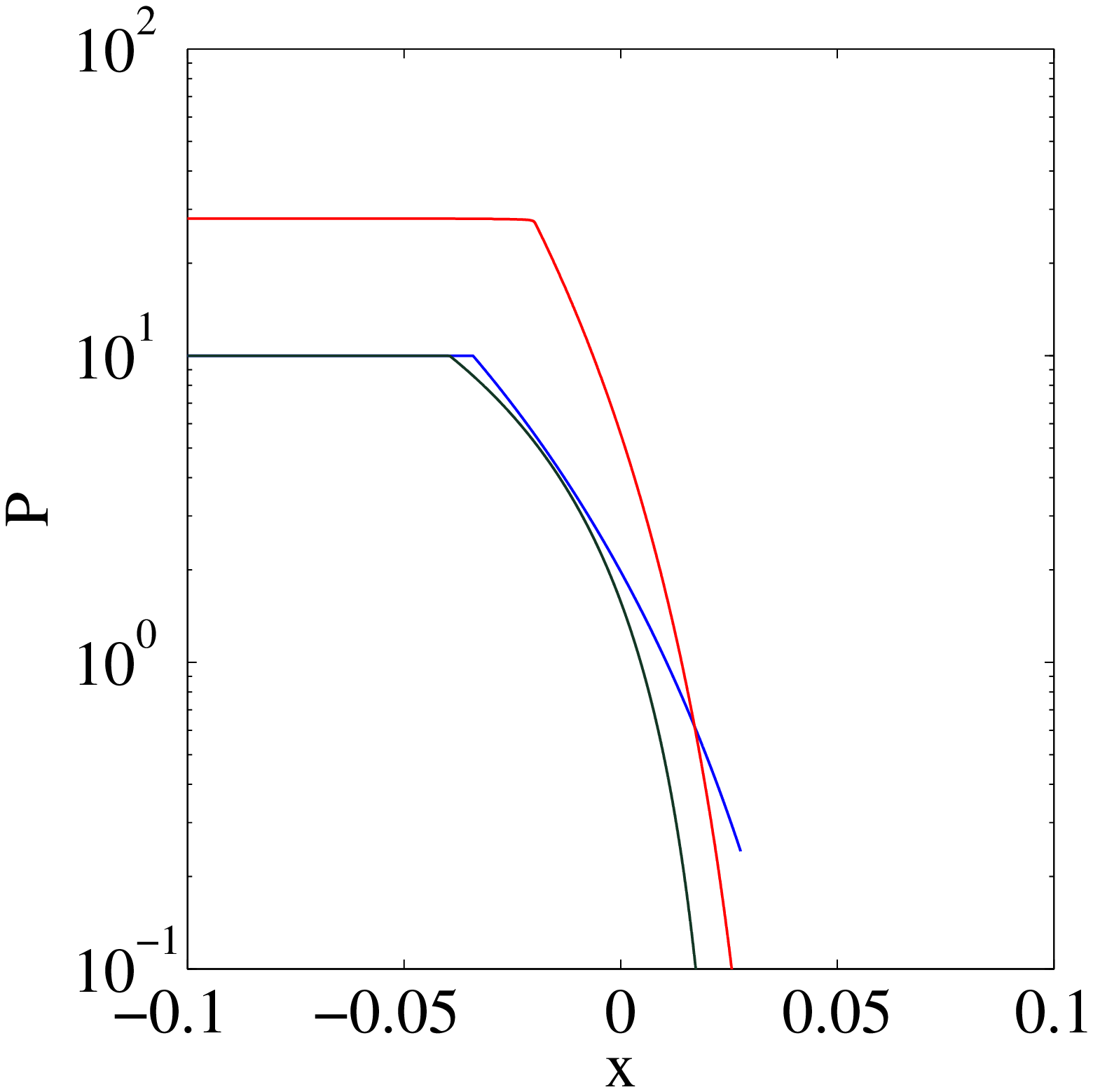}
    \includegraphics[width=0.32\textwidth,angle=0]{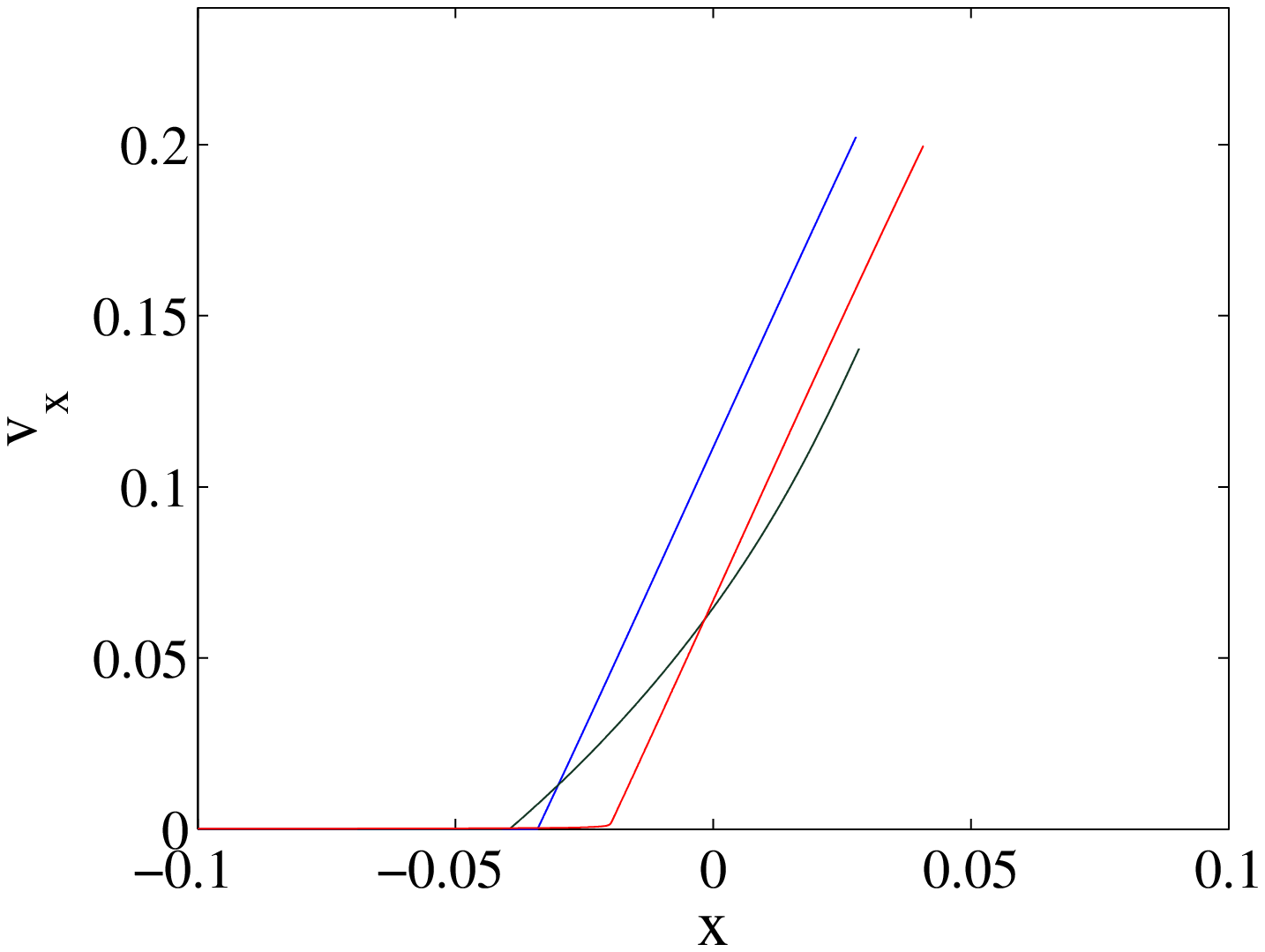}
    \includegraphics[width=0.31\textwidth,angle=0]{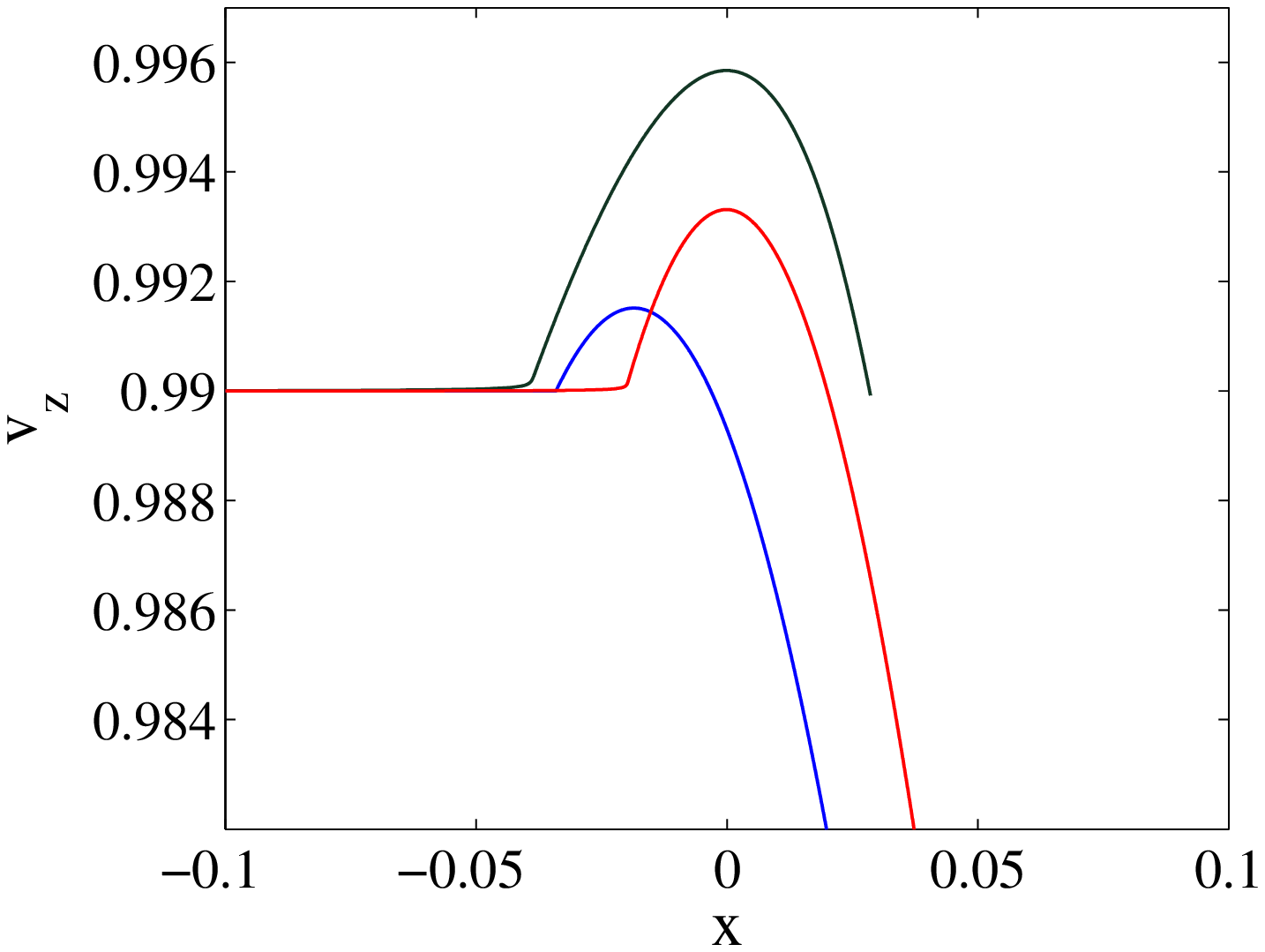}
    \caption{The three models corresponding to the numerical simulations cited in the main text (MHDA blue, MHDB green, HDB red).}
\label{figMiz}
\end{figure*}

As a final application we examine the consistency of our steady-state solution with the simulations appearing in \cite{Mizuno_2008}, and their similar ones in \cite{Zenitani_2010}. Using the same set of initial parameters (Table~\ref{arrayinitialMiz}\footnote{Note that the values of ``magnetic fields'' given in \cite{Mizuno_2008} are in fact the magnetic field over $\sqrt{4 \pi}$.}), the results obtained (Fig.~\ref{figMiz}) are in excellent agreement with the simulations. The situation is identical until of course the point, where a contact discontinuity occurs, the left plateau at their diagrams, corresponding to a termination of the rarefaction process. This picture is expected whenever a nonzero external pressure/density exists, since a contact discontinuity between the two fluids, but also a shocked region in the exterior medium are formed, see next section.

\section{Interpretation of the results - Discussion}
\label{discussion}

The main characteristic of the rarefaction model presented above is the significant acceleration of the flow. Depending on the available energy reservoir this acceleration is performed in expense of the Poynting energy ($\enth \lrz\sigma$, magnetically driven), of the thermal one ($\enth$, thermally driven), or both (mixed type). But no mater in what form the leading energy is, in all of the flows the acceleration achieved leads to completely matter-dominated flows ($\lrz_{max}\sim \mu$).

A phenomenological interpretation for the acceleration spatial scale is based on the magnetization parameter power law ($\sigma \propto r^{-2/3}$) that applies to all cold flows, even if small deviations like in (EP) model exist. Since in all cases the same initial values of $\mu,\lrz,\sigma$ apply, the relative distances are ascribed to the early appearance of the rarefaction process.  In contrast to this, in the thermal driven rarefaction (TD) the acceleration takes place in much greater distances, despite the early appearance of the rarefaction wave front. The lengthier, and thus less efficient action, is associated with the conversion of the thermal energy to bulk kinetic, which follows a much shallower law than the Poynting energy decrease (see Fig.~\ref{figrestmods}). For the case of a mixed type rarefaction where the poloidal magnetic field is significant, ($B_p>|B_y|$) and $\belpa^2 \sim 1 - \epsilon$ with $\epsilon<1/\enth\lrz$. Thus $\sigma \propto 1/M^2 \propto \dens/\enth \propto \dens^{2-\polind}$, while if significant thermal context exists, then $\enth \propto \dens^{\polind-1}$ with $\polind=4/3$ exhibiting the slower scale of the thermal conversion. That result is in agreement with other models where rarefaction was considered in the negligible poloidal magnetic field limit, in both planar \cite{Kostas_2013} and axisymmetric flows \cite{Komissarov_Vlahakis_Konigl_2010}.

The rarefaction wave front is determined by the vanishing of the denominator of Eq.~(\ref{denom}). In general that vanishing occurs at the modified fast magnetosonic surface which corresponds to points where $\lrz \theta /c$ equals the comoving phase velocity of the fast magnetosonic waves. Thus the lines of $\theta=$const are also the characteristics of our system, exactly as in the hydrodynamic homogeneous rarefaction, forming around the poloidal velocity the Mach cone of the fast magnetosonic velocities; Fig.~2 in \cite{Kostas_2013} is very instructive. Considering the cone at the axis origin the wave front is the envelope surface of all the fast magnetosonic disturbances emitted from the point of the boundary discontinuity. As such, the presence of a poloidal magnetic component and transverse initial velocity both affect the wave front inclination and $\theta_{RW}$ is no longer given by a simple expression like $\sqrt{\sigma_0}/\lrz_0$ as in the $B_p \ll |B_y|$, $v_{y0}\approx 0$ flows, but from the more complicated expression~(\ref{eqnassympPM}) and Fig.~\ref{thapp}. For the mixed type scenario (TD), the Mach cone is not obtained by the sonic disturbances expected to propagate in a much narrower opening of $u_s/(\lrz_0 v_{p0}) \sim 6.5 \times 10^{-3}$, if the flow was purely hydrodynamic, but from the more extended one of the fast magnetosonic disturbances ($c_s<c_f$, where $c_f$ the velocity of the fast magnetosonic disturbances at the comoving frame, which is significantly affected by the presence of $B_p$).

The cold and uniform flow is studied in detail in Appendix~\ref{appA}, where the scaling of $M^2 \propto r^{-2/3}$ is formally derived. Beyond that, the same law can be derived by more intuitive arguments. The invariance of
\begin{equation}
\label{eqBcscal}
    \left(\lrz v_p\right)^2 \frac{B^2-E^2}{ B_p^2}=\mbox{const}
    \quad \Leftrightarrow \quad \frac{B_{\rm co}}{\dens}=\mbox{const} \,,
\end{equation}
where $B_{\rm co}$ the magnetic field in the comoving frame, is connected to the magnetic flux conservation and is obtained using the integral expressions.\footnote{In the general case of nonnegligible thermal content we find $\left(\lrz v_p/c\right)^2 (B^2-E^2)/B_p^2=\belpot^2-1+\mu^2(1-\belpa^2)^2/\enth^2$. It is interesting to note that the same relation holds in the axisymmetric case as well, with $\belpot$ however replaced by $E/B_p=\varpi\Omega/c$, the cylindrical distance in units of the light cylinder radius.}
The continuity equation and the flux conservation along the streamline provides $\dens\lrz v_\theta=\massint (\nbl\bflux)_r \sim \massint \bflux/r$, which is used in conjunction with $D=0\Leftrightarrow (\lrz v_\theta)^2 = (B^2-E^2)/(4\pi\dens)$ (Eq.~\ref{denom}) to obtain $B^2-E^2 \propto 1/(\dens r^2)$. Combining this scale with Eq.~(\ref{eqBcscal}), we derive the density evolution $\dens \propto r^{-2/3}$ and by Eq.~(\ref{Mdef}) the requested $M^2\propto r^{2/3}$ power law.

The transverse velocity evolution induces implications in some of the models considered (TD, EP, LP03), while both cases of an increasing (LP models) or a decreasing profile of $v_y$ exist. The general behavior of both velocity components is obtained in terms of $\sigma$ in the analysis of Appendix~\ref{appA}, see Eqs.~(\ref{uyanal}), (\ref{upanal}), according to which $\belpa^2>1$ leads to the decrease, while $\belpa^2<1$ to the increase of the transverse velocity\footnote{In an axisymmetric trans-Alfv\'enic flow $\belpa^2<1$ always hold since at the Alfv\'en surface $M^2=1-\belpa^2$. In a planar symmetric flow this limitation does not exist, and thus we included cases with $\belpa^2>1$ in our study.}. The asymptotic values of the velocity components is obtained by setting $\sigma \to 0$
\begin{equation}
    \frac{v_{y\infty}}{c}=\frac{\belpa^2}{\belpot}\,, \quad \frac{v_{p\infty}}{c}=\frac{v_{r\infty}}{c}=\sqrt{1-\frac{\belpa^4}{\belpot^2}-\frac{1}{\mu^2}} \,,
\end{equation}
and simply state that all the plasma momentum in the $y$-direction has been transferred to the matter. The momentum conservation dictates the increase of the matter's transverse momentum ($\lrz v_y$), but the acceleration at the poloidal direction leads also to the inertial matter enhancement ($\lrz$). Weather inertial increase suppresses the transverse velocity one, or not, is not uniquely determined; it depends on the initial conditions in the way that the relationship above determines.
The velocity expressions are also useful in the calculation of the magnetic components asymptotic ratio. For that purpose, we apply their value to the $\belpot$ integral, and after some manipulation we find
\begin{equation}
    \frac{\ratb_\infty}{\ratb_0}=\frac{v_{p0}}{v_{p\infty}}\frac{M_0^2+\belpa^2-1}{M_0^2}
\end{equation}
by which the insignificant alteration of the $\ratb$ ratio is concluded.

The evolution of the remaining parameters can also be understood analytically. The transverse magnetic field is easily obtained by Eq.~(\ref{eqnsby}), which reveals the $B_y \propto r^{-2/3}$ decrease, but at distances where $M^2\gg \belpot^2$ due to the presence of the $\belpot^2$ term in the denominator. The $\ratb$ evolution indicates then that the poloidal component follows the same evolution $B_p \propto r^{-2/3}$, except for the case where a minor deviation of $\ratb$ is observed. It is instructive to compare the scaling of the magnetic field component with the one obtained by the axissymetric MHD steady-state models, $B_p\propto 1/\varpi^2$, $B_\phi\propto 1/\varpi$, see for example \cite{Vlahakis_2003a}. Besides the differences in the decrease of the two components, in rarefaction the Poynting energy conversion $\mu-\enth \lrz \propto |B_y| \propto r^{-2/3}$ is to be compared with the one obtained in the semi-analytical results of \cite{Vlahakis_2003a} and the numerical ones (\cite{Komissarov_Vlahakis_Barkov_2009} and references therein) where the conversion is much slower caused by the slow decrease of $\varpi B_\phi$. Thus the magnetic driven rarefaction is to be considered as a powerful and short scaling mechanism to convert Poynting energy to bulk kinetic, and as such its contribution to the high energy astrophysical phenomena might be important.

The magnetization parameter does follow the $-2/3$ power law in a great extend, but as Eq.~(\ref{usfleqM2}) suggests, this behavior deviates whenever $\ratb$ is close to unity and $v_y \sim \belpot$, e.g. in (EP) model. The deviation has not serious implications in the scalings of most physical quantities, but it affects the calculations for the maximum extend of the rarefied flow ($\theta_{PM}$), see Eq.~(\ref{eqnassympPM}). The two terms appearing in this expression are of the same order magnitude and thus the accurate numerical integration is unavoidable, especially for low $\ratb$ flows. The resulting $\theta_{PM}$ for low initial transverse velocity as a function of $\ratb_0$ are shown in Fig.~\ref{thapp}.

Our model describes the rarefaction until its full completion, leading to a completely matter-dominated flow that fills the space up to polar angle $\theta_{PM}$. This will be indeed the end state if the flow is surrounded by void space. If the pressure or density of the environment is nonzero then the expansion of the flow will modify the environment as well, and a contact discontinuity (CD) will be formed. The details of the final state depend on the characteristics of the environment (for example if it is a hydrodynamic super-sonic flow in the $\hat z$ direction a shocked region will be formed and the pressure at the CD will be related to the shock jump conditions). For an initially uniform environment the CD will be conical $\theta=\theta_{CD}$ passing trough the origin. The environment characteristics define the value of the pressure $P_{CD}$ at CD, and thus the rarefaction ceases at some angle $\theta_F$ in which pressure equilibrium is reached
\begin{equation}
\label{nonzeroext}
\left[\frac{B^2-E^2}{8\pi} + P\right]_{\theta=\theta_F} = P_{CD}\,.
\end{equation}
The above equation defines the angle $\theta_F$, after which (and up to $\theta_{CD}$) the flow remains uniform. Our model correctly describes the rarefaction till the flow becomes uniform, so we can use it to find the end state from Eq.~(\ref{nonzeroext}), and also find $\theta_{CD}=\left[\vartheta\right]_{\theta=\theta_F}$.

If the flow is cold then Eq.~(\ref{nonzeroext}) can be much simplified. Since the comoving magnetic field
scales with the density $\sqrt{B^2-E^2} \propto \rho \propto 1/M^2$, Eq.~(\ref{nonzeroext}) can be rewritten as
\begin{equation}
\label{nonzeroextcold}
\left[\frac{M_0^2}{M^2}\right]_{\theta=\theta_F} = \sqrt{\frac{8\pi P_{CD}}{B_0^2-E_0^2}}\,.
\end{equation}
If for example the environment is a uniform static medium with pressure 25 times smaller than the initial magnetic pressure of the flow, then at $\theta_F$ the ratio $M_0^2/M^2=0.2$. Using the third row of Fig.~\ref{figcoldmods}, e.g. for model LP, we find $r/r_0\approx 300 $, and from the other diagrams for the same model all the rest physical quantities. The terminal Lorentz factor is $\sim 350 $ (corresponding to efficiency $\lrz/\mu \sim 30\%$) and the flow inclination $\vartheta=\theta_{CD}=0.5 \vartheta_\infty=0.16^\circ$.

More complicated environments are beyond the scope of this paper, but they are definitely an interesting application of the model. Possibly the environment itself can also be modeled with an $r$ self-similar model. In this paper we focus on the strongly magnetized cases and highly relativistic velocities, however the model applies to other cases as well, for example to slow magnetosonic weak discontinuities; these will be examined in another connection.

\begin{acknowledgments}
This research has been co-financed by the European Union (European Social Fund -- ESF) and Greek national funds through the Operational Program ``Education and Lifelong Learning'' of the National Strategic Reference Framework (NSRF) - Research Funding Program: Heracleitus II. Investing in knowledge society through the European Social Fund.
\end{acknowledgments}

\appendix

\section{The cold homogeneous case}
\label{appA}
We derive here the analytical expressions for the cold ($\enth \to 1, u_s \to 0$)) and homogeneous flow ($\modind=1$, see Eq.~\ref{eqnsdens}), that serves as an extension of the hydrodynamic solutions of \cite{Landau_fluid} and \cite{Granik_1982}. At that limit, the rarefaction wave front is determined straightforwardly by the vanishing of the denominator of Eq.~(\ref{Meq}). The qualitative behavior of the solution is easily understood, if we use Eq.~(\ref{denom}):
\begin{equation}
\label{denomcold}
    D=0 \Rightarrow \; \left(\frac{\lrz v_\theta}{c}\right)^2=\frac{B^2-E^2}{4\pi\dens c^2} \,.
\end{equation}
As we proceed to the integration from the initial surface $\theta_0 = -\pi/2$ to the higher angle ones, the $d/d\theta$ derivatives equal to zero indicating the uniform flow phase. That uniformity breaks at the point $\theta_{RW}$ where the $\hat\theta-$component of the flow proper velocity becomes equal to the fast magnetosonic one yielding the weak discontinuity. From this angle and beyond a (0/0) form arises and the derivatives attain a finite value signaling the initiation of the rarefaction process.

It is easy to derive an analytical expression for the rarefaction wave angle in terms of the initial quantities and not only for the cold limit. For this purpose, we combine the vanishing of $D$ (Eq.~\ref{denom}) and the Bernoulli equation~(\ref{bernoulli}) to eliminate $f$ and find
\begin{eqnarray}
\label{rwanglenoassump}
\sin ^2\left(\vartheta-\theta\right)= \frac{\sigma ^2 M^2}{\belpot^2 \left(v_p/c\right)^2}-\frac{\belpot^2-1}{M^2}
\nonumber  \quad \\
- \frac{1-M^2-\belpot^2}{M^2 \left(\lrz v_p/c\right)^2}u_s^2 \,, \quad
\end{eqnarray}
where use of Eq.~(\ref{eqnsdens})-(\ref{eqnsgamvy}) was also made. We would like to underline that the above expression provides not only the rarefaction wave front angle $\theta_{RW}$, i.e. when we consider the initial values of the quantities, but also relates the appearing quantities at the subsequent rarefaction phase.

A point of special attention for the following calculations is the transverse velocity $v_y$ which, even if it is negligible at the beginning, it is possible to end up with significant values (see for example models EP, TD) affecting the derived asymptotic expressions. Under this perspective two helpful and accurate expressions are:
\begin{eqnarray}
\label{usfleqM2}
M^2=\frac{\belpot}{\sigma}\left(\belpot-\frac{v_y}{c}\right) \,, \\
\label{usfleqxa2}
\belpa^2=\frac{\enth \lrz \belpot }{\mu}\left(\frac{v_y}{c}+\frac{\sigma}{\belpot}\right) \,.
\end{eqnarray}
It is also useful to express the velocity components in terms of the magnetization parameter. Eq.~(\ref{eqnsgamma}), (\ref{eqnsgamvy}), (\ref{eqnssigM}) yield
\begin{eqnarray}
\label{uyanal}
    \frac{v_y}{c}=\frac{\belpa^2+\sigma\left(\belpa^2-1\right)}{\belpot}\,, \quad \\
\label{upanal}
    \frac{v_p}{c}=\sqrt{1-\frac{\belpa^4}{\belpot^2}\left(1+\frac{\belpa^2-1}{\belpa^2}\sigma\right)^2-\frac{\enth^2}{\mu^2}\left(1+\sigma\right)^2} \,. \quad
\end{eqnarray}

\begin{figure}
    \includegraphics[width=0.35\textwidth,angle=0]{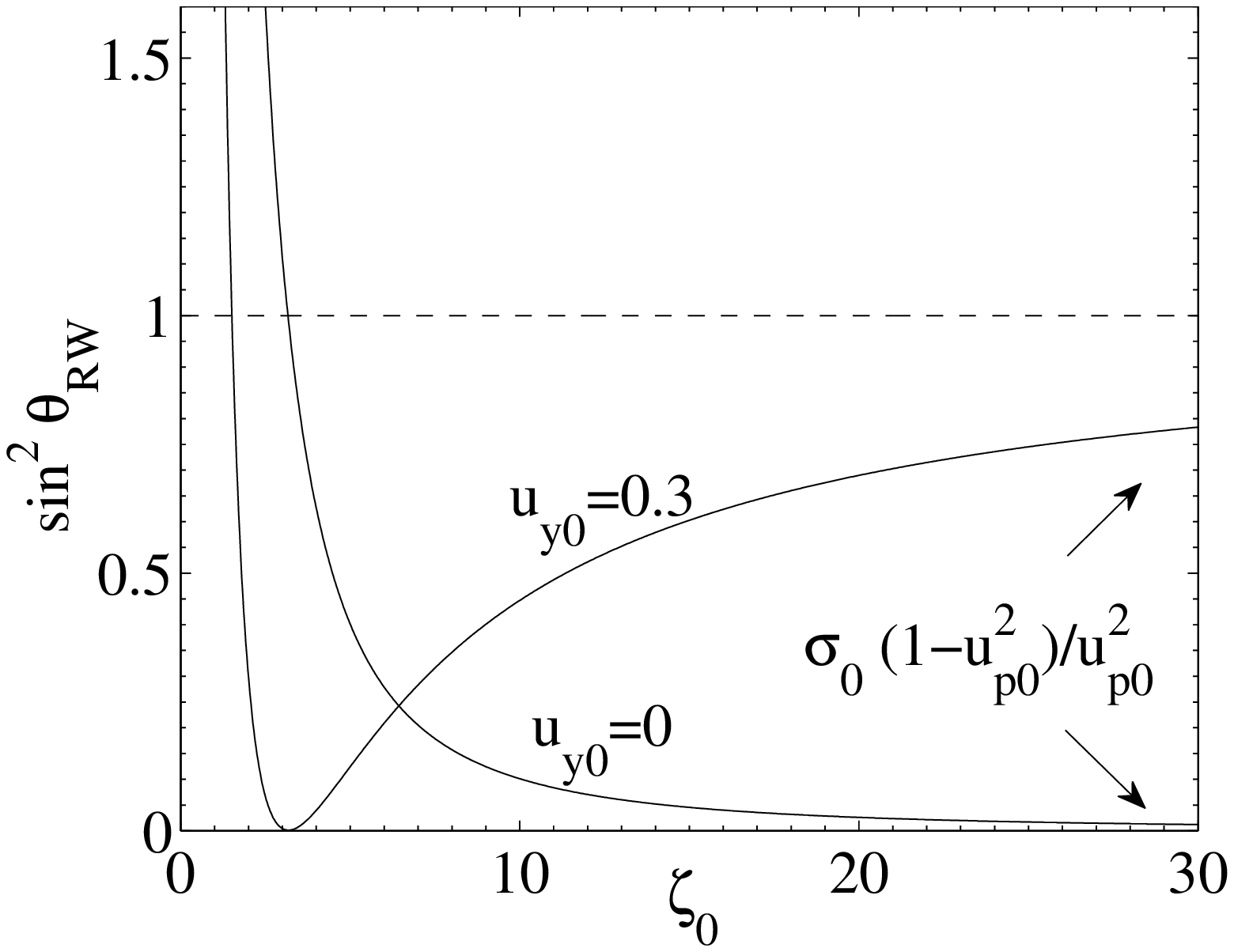}\\
    \vspace{0.2cm}
    \includegraphics[width=0.35\textwidth,angle=0]{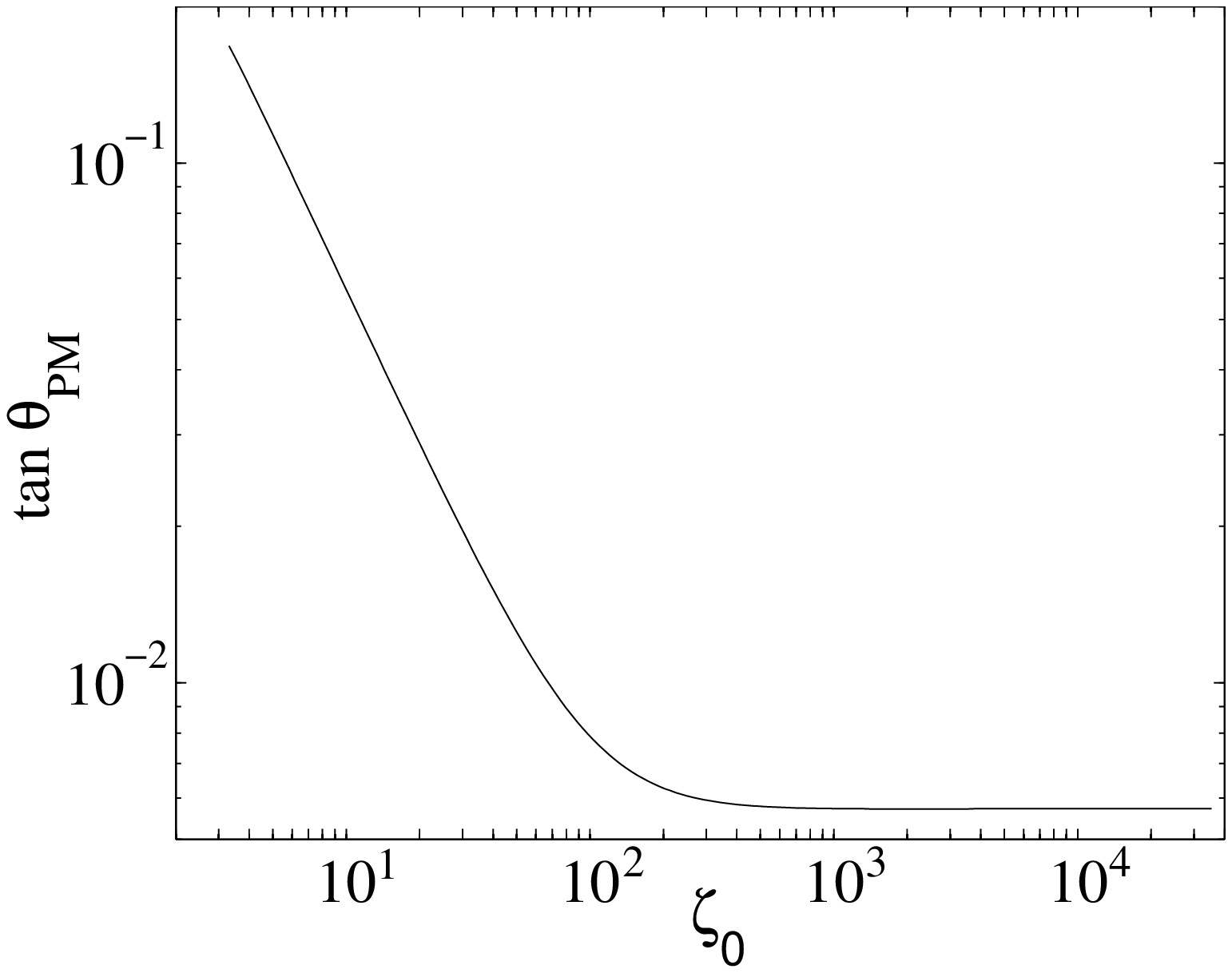}
    \caption{The calculated angles for a cold flow ($\lrz_0=100$, $\sigma_0=10$) as a function of $\ratb_0$. \textbf{Top:} The rarefaction front ($\sin^2\theta_{RW}$) for two different values of the transverse velocity, and the asymptotic expression ($\ratb_0 \to\infty$). Notice that for improper initial conditions (low $\ratb_0$ or high $v_{y0}$) we obtain $\sin^2\theta_{RW}>1$ corresponding to sub-fast magnetosonic flow where the rarefaction wave is impossible. \textbf{Bottom:} The resulting extension of the rarefied area ($\tan \theta_{PM}$) for a cold flow of negligible $v_{y0}$.}
\label{thapp}
\end{figure}

We now focus on the cold flow limit and we use Eq.~(\ref{usfleqM2}) in Eq.~(\ref{rwanglenoassump}) to obtain
\begin{equation}
\label{rwanglenoassump2}
\sin^2\left(\vartheta-\theta\right)=\frac{\sigma}{\belpot}\left[\frac{\belpot- \frac{v_y}{c}}{\left(v_p /c\right)^2}+\frac{1-\belpot^2}{\belpot-\frac{v_y}{c}}\right] \,.
\end{equation}
The effects of the transverse velocity are important in cases where the integral $\belpot$ is close to unity, i.e. when the ratio $\ratb$ is comparable to the transverse velocity. Assuming the initial values we obtain the rarefaction wave front angle; for a specific $\lrz_0$, the angle depends on both $v_{y0},\,\ratb_0$ via $\belpot$. In Fig.~\ref{thapp} we give the relevant plot as function of $\ratb_0$ for two different values of the initial transverse velocity. Notice that at the limit of the negligible poloidal magnetic field ($B_p \to 0, \, \belpot \to \infty$) the above expression becomes $\sin^2\theta_{RW}=\sigma_0 (c^2-v_{p0}^2)/v_{p0}^2$ in agreement with the results of \cite{Kostas_2013}.

In order to derive the spatial evolution of the integrable quantities we use Eqs.~(\ref{bernoulli}), (\ref{denom}) to eliminate $\vartheta$ this time:
\begin{eqnarray}
\label{appassymptotnoassump}
\frac{1}{f^2} = \frac{\belpot^2 \enth^2 \lrz^2 \sigma^2}{M^2}
-\belpot^4\enth^2\frac{\belpot^2-1}{M^6}\left(\lrz\frac{v_p}{c}\right)^2 \quad \nonumber \\
-\belpot^4\enth^2\frac{1-M^2-\belpot^2}{M^6}u_s^2 \,. \quad
\end{eqnarray}
In the cold limit and by use of Eqs.~(\ref{usfleqM2}), (\ref{usfleqxa2}), (\ref{uyanal}), (\ref{eqnsgamma}) we obtain a rather simple expression
\begin{equation}
\label{appasymM2}
    \frac{1}{f^2}=\frac{\belpot^4}{M^6} \left[ \left(1-\belpa^2\right)^2 \mu^2+\belpot^2-1 \right] \,,
\end{equation}
which gives the power law evolution of the Alfv\'enic Mach number $M^2 \propto f ^{2/3}\propto r ^{2/3}$. It is instructive to compare this result with the one obtained at the negligible poloidal case. In that limit both $M^2$ and $\belpot^2$ become infinite, but their ratio retains the finite value of $1/\sigma$, see Eq.~(\ref{eqnssigM}). Thus the same spatial scaling is provided in terms of $\sigma \propto r^{-2/3}$.

We use this simple result to calculate the maximum extend of the rarefied area. For that purpose Eq.~(\ref{feq}) provides
\begin{eqnarray}
\frac{d\theta}{d M^2}=\frac{3}{2}\frac{\tan\left(\vartheta-\theta\right)}{M^2} \,.
\end{eqnarray}
The tangent appearing is obtained from the Bernoulli Eq.~(\ref{bernoulli})
\begin{equation}
\label{eqappsin}
    \sin\left(\vartheta-\theta\right)=\frac{M^2 + \belpot^2-1}{\sqrt{M^2 \ H\left(M^2\right)}} \,,
\end{equation}
where $H(M^2)$ is a polynomial of $M^2$
\begin{eqnarray*}
    H\left(M^2\right)=\frac{\left[\mu^2\left(\belpot^2-\belpa^4\right)-\belpot^2\right] M^4}{\left[\mu^2(1-\belpa^2)^2+\belpot^2-1\right]\belpot^2}-2M^2+1-\belpot^2 \,.
\end{eqnarray*}
The resulting expression must be calculated numerically
\begin{eqnarray}
\label{eqnassympPM}
    \theta_{PM}=\theta_{RW}+ \qquad \qquad \qquad \qquad \qquad \qquad \qquad \nonumber\\
    +\displaystyle\frac{3}{2}\int^\infty_{M^2_0} \! \frac{M^2 +\belpot^2-1}{\sqrt{M^2 H(M^2)-\left(M^2+\belpot^2-1\right)^2}} \frac{dM^2}{M^2} \,. \quad
\end{eqnarray}
In Fig.~\ref{thapp} we show the obtained $\theta_{PM}$ values as a function of the $\ratb_0$ for the same model parameters ($\modind,\mu,\lrz_0,v_{y0}$) like the ones used in (LP, MP, EP) models..

In the limit of negligible poloidal magnetic field and $v_y=0$ we get $H(M^2)=\sigma^{-1} M^2[\mu^2-(\sigma+1)^2]$; the second term is of order $\sim \mu^2/\lrz^2$ and thus can be ignored. The transformation $dM^2 /M^2 = -d\sigma /\sigma $ provides
\begin{eqnarray*}
    \theta_{PM}=\theta_{RW}+\displaystyle\frac{3}{2}\int_0^{\sigma_0} \! \frac{1+\sigma}{\mu \sqrt \sigma}\,\, d\sigma \,.
\end{eqnarray*}
In that limit, $\theta_{RW}\approx -\sqrt\sigma_0/\lrz_0$, and the above calculation yields $\theta_{PM}=2\sqrt\sigma / \mu$ which is exactly the result found in \cite{Kostas_2013}.

\bibliographystyle{aipauth4-1}
\bibliography{Kostas}

\end{document}